\documentclass[trackchanges, twocolumn]{aastex701}

\usepackage{graphicx}
\usepackage{grffile}
\usepackage{amsmath}
\usepackage{amssymb}
\usepackage{bm}
\usepackage{hyperref}
\usepackage{makecell}
\usepackage{booktabs}   
\usepackage{longtable}   
\usepackage{threeparttable} 
\usepackage{comment}
\usepackage{placeins}
\usepackage{orcidlink}  
\usepackage{enumitem}
\usepackage[normalem]{ulem}
\usepackage{booktabs}
\usepackage{tabularx}
\usepackage{pifont}   
\usepackage{siunitx}  
\usepackage{array}
\usepackage{xcolor}
\usepackage{rotating}

\usepackage{enumitem}
\usepackage{needspace}

\setlist[enumerate,1]{leftmargin=*}

\newcommand{\music}{\textsc{music2}}
\newcommand{\CG}{\textsc{CosmicGrain}~}
\newcommand{\gadget}{\textsc{Gadget-4}~}

\begin{document}

\title{CosmicGrain: An On-the-fly Dust Physics Module}

\author[orcid=0009-0004-8503-0483]{Ezra Huscher}
\affiliation{New Mexico State University, Department of Astronomy}
\affiliation{Cosmic Dawn Center (DAWN) Niels Bohr Institute, University of Copenhagen, Denmark}
\email[show]{ez@nmsu.edu}  

\author[orcid=0000-0002-0496-1656]{Kristian Finlator} 
\affiliation{New Mexico State University, Department of Astronomy}
\affiliation{Cosmic Dawn Center (DAWN) Niels Bohr Institute, University of Copenhagen, Denmark}
\email{finlator@nmsu.edu}

\begin{abstract}
We present \textsc{CosmicGrain}, a dust physics module built on \gadget that models explicit dust superparticles that experience resolved gas-dust drag and radiation pressure, undergo collisional heating and cooling with gas, and are further shaped by subgrid treatments for grain growth, thermal sputtering, coagulation, shattering, supernova shock destruction, and astration. We run an eleven-rung physics ladder (S0--S10) at $1024^3$ resolution to isolate the contribution of individual processes to a Milky Way-mass halo's dust population. Because dust is evolved as active on-the-fly particles, individual grains retain distinct histories producing results inaccessible to passive dust models such as the strongly asymmetric silicate-to-carbon shattering destruction ratio driven by silicate grains' smaller birth size. Gas-dust drag is the dominant driver of ISM dust enrichment, increasing dust levels by $\sim100\times$ while every other process changes the $R_{200}$ dust abundance by only 10--30\%. We find reasonable agreement with the Milky Way's radial dust-to-gas and dust-to-metal ratios, with carbon dust preferentially enriching the CGM relative to silicates. The code is freely available at \url{https://cosmicgrain.space}.


\end{abstract}


\section{Introduction}
\label{sec:intro}

Cosmic dust is a prominent factor in determining the physical and observable properties of galaxies. It can store and release metals, influence gas cooling and heating, and change the speed and strength of gas falling in or getting pushed out of galaxies, all of which will influence the formation and composition of new stars. Broadly, dust describes condensations of free-floating metals that can increase or decrease in size, continually be created, continually be destroyed, have its composition recycled, absorb or emit energy, and be propelled around space. Modeling dust is challenging due to this diversity of simultaneous processes.

In light of recent discoveries of unexpectedly dusty galaxies at high redshift \citep{zavala23, gottumukkala24}, discrepancies between dust models and observations suggest that existing frameworks may be missing important physics \citep{parente25}. Part of this gap may involve how dust itself is represented in cosmological simulations which is broadly done in one of three ways: 1.) as a passive scalar advected with gas \citep{vogelsberger20, dave19, k22}, 2.) coupled to gas but with an approximate grain size distribution, such as the widely-used two-size approximation \citep{hirashita15, aoyama17}, 3.) and as independent live particles subject to their own behavior \citep{bekki15, mckinnon18}. Despite this range of sophistication, a near-universal assumption is that dust and gas share a velocity field — even live-particle experiments in isolated disc simulations concluded that strong coupling remains a reasonable approximation at cosmological resolution \citep{mckinnon18}. \citet{li21} pushed the live-particle approach furthest to date, evolving 41 grain size bins in a cosmological volume, but in a small 12Mpc box at a modest $256^3$ resolution.

The complex physical processes involved in dust evolution can be modeled most accurately with cosmological zoom-in simulations which provide improved resolution in the interstellar medium (ISM). Many of these processes require cold temperatures and high densities which large volume simulations cannot achieve therefore creating more reliance on subgrid models \citep{trayford26}. An ideal dust model, while always contending with the limitations of resolution and computation, will employ explicit and independent dust particles, grain size tracking as a continuous range, and a complete physics ladder from creation to full physics which isolates each process's contribution to established observables. To our knowledge, no existing cosmological simulation has incorporated all three of these elements together. This is the goal of this work.

\textsc{CosmicGrain} is a cosmological zoom simulation modeling a single Milky Way-mass halo treating dust grains as explicit N-body superparticles with independent kinematics, drag, gravity, and radiation pressure — extending the live-particle dust treatment introduced by \citealt{bekki15} and \citealt{mckinnon18} in isolated-galaxy simulations to a fully cosmological zoom context. This cosmological setting exposes the model to a circumgalactic medium and a merger history that the halo builds up over cosmic time, environmental context that is absent by design in isolated-galaxy simulations.

Each superparticle carries a single representative grain radius that evolves freely as interactions with gas, stars, and other dust impart changes, with the ensemble of superparticles across the simulation volume collectively sampling the galaxy's grain size distribution. The complete dust model is run with 3 resolutions ($512^3$, $1024^3$, $2048^3$) to explore convergence and applied across an incremental physics ladder ($1024^3$) to examine direct impacts of isolated dust processes on observables such as the Dust-to-Gas (D/G) and Dust-to-Metals (D/Z) ratios.

In Section \ref{sec:methods}, we discuss the formulation and details of the simulation. In Section \ref{sec:dust_processes}, we examine all of the dust processes one-by-one and their implementation in detail. In Section \ref{sec:results}, we present the simulation results. Lastly in Section \ref{sec:discussion}, we discuss the comparison to observations and other models as well as limitations and future improvements.

\begin{table*}
\centering
\caption{%
    Dust physics simulation grid. Each row adds one physical process to the previous configuration. Columns indicate which processes are active (\checkmark) or disabled (--) for each simulation label.
    All runs use the same \gadget binary and \texttt{Config.sh} while dust processes are toggled via the \texttt{DustEnable*} switches in the parameter file.
}
\label{tab:sim_grid}
\setlength{\tabcolsep}{3pt}
\begin{tabular}{l l ccccccccccc ccc}
\toprule
 &  &
\multicolumn{11}{c}{Dust Processes} &
\multicolumn{3}{c}{Resolution} \\
\cmidrule(lr){3-13} \cmidrule(lr){14-16}
Label & Description &
\rotatebox{70}{Creation} &
\rotatebox{70}{Cooling} &
\rotatebox{70}{Drag} &
\rotatebox{70}{Astration} &
\rotatebox{70}{Sputtering} &
\rotatebox{70}{Growth} &
\rotatebox{70}{Clumping} &
\rotatebox{70}{Shock destr.} &
\rotatebox{70}{Coagulation} &
\rotatebox{70}{Shattering} &
\rotatebox{70}{Rad. pressure} &
\rotatebox{70}{$512^3$} &
\rotatebox{70}{$1024^3$} &
\rotatebox{70}{$2048^3$} \\
\midrule
S0  & Creation only (baseline)         & \checkmark & -- & -- & -- & -- & -- & -- & -- & -- & -- & -- & --  & \checkmark & -- \\
S1  & $+$ Cooling                            & \checkmark & \checkmark & -- & -- & -- & -- & -- & -- & -- & --  & --& -- & \checkmark & -- \\
S2  & $+$ Drag                       & \checkmark & \checkmark & \checkmark & -- & -- & -- & -- & -- & -- & --  & --& -- & \checkmark & -- \\
S3  & $+$ Astration                          & \checkmark & \checkmark & \checkmark & \checkmark & -- & -- & -- & -- & -- & --  & --& -- & \checkmark & -- \\
S4  & $+$ Thermal sputtering                 & \checkmark & \checkmark & \checkmark & \checkmark & \checkmark & -- & -- & -- & -- & --  & --& -- & \checkmark & -- \\
S5  & $+$ Grain growth                       & \checkmark & \checkmark & \checkmark & \checkmark & \checkmark & \checkmark & -- & -- & -- & --  & --& -- & \checkmark & -- \\
S6  & $+$ Subgrid clumping                   & \checkmark & \checkmark & \checkmark & \checkmark & \checkmark & \checkmark & \checkmark & -- & -- & -- & --  & --& \checkmark & --  \\
S7  & $+$ SN shock destruction               & \checkmark & \checkmark & \checkmark & \checkmark & \checkmark & \checkmark & \checkmark & \checkmark & -- & -- & --  & --& \checkmark & -- \\
S8  & $+$ Coagulation                        & \checkmark & \checkmark & \checkmark & \checkmark & \checkmark & \checkmark & \checkmark & \checkmark & \checkmark & -- & -- & -- & \checkmark & -- \\
S9  & $+$ Shattering    & \checkmark & \checkmark & \checkmark & \checkmark & \checkmark & \checkmark & \checkmark & \checkmark & \checkmark & \checkmark & --  & -- & \checkmark & -- \\
S10 & $+$ Radiation pressure (full)  & \checkmark & \checkmark & \checkmark & \checkmark & \checkmark & \checkmark & \checkmark & \checkmark & \checkmark & \checkmark & \checkmark & \checkmark & \checkmark & \checkmark \\
\bottomrule
\end{tabular}
\end{table*}

\section{Methods}
\label{sec:methods}
\subsection{Initial Conditions with CAMB and MUSIC}

To set up initial conditions for our zoom simulation, we generated linear transfer functions with the code CAMB \citep{lewischallinor11} for the Planck 2018 cosmology \citep{planck18} results where $\Omega_m$ = 0.3158, $\Omega_L$ = 0.6842, $\Omega_b$ = 0.04936, and $H_0$ = 0.6732. The transfer function file tabulates density and velocity transfer functions for dark matter and baryons, as functions of the comoving wavenumber k/h. MUSIC2, an update to the robust generator of cosmological initial conditions from \citealt{hahn11}, uses these transfer functions to produce primordial Gaussian density and velocity fields.

We initially run a 50 Mpc$^3$ $h^{-1}$ volume with just low-resolution dark matter to select our desired high-resolution region centered on a Milky Way mass halo. We then re-run the simulation with nested zoom regions of increasing resolution to temper the resolution transitions.

We set the \music~initial conditions to level 10 in the zoom region corresponding to a resolution equivalent of $1024^3$ particles. Our resulting 50 Mpc box contains 2,128,487 low-resolution dark matter particles and is centered on a high-resolution region about 7 Mpc across that contains one Milky Way size halo and numerous smaller satellite galaxies. 

\subsection{Simulation}
\label{sec:simulation}

We cloned \gadget \citep{springel21} and use the  star formation model of \citealt{springel03} with its effective equation of state, but replace the implicit supernova feedback with an explicit thermal feedback scheme that deposits energy directly to neighboring gas particles.

Our dust model was designed to be as flexible as possible, with extensive tunable parameters (see Table \ref{tab:new_gadget4_params}) and built-in adaptation to handle different resolutions. That is, where possible, we implement in \CG resolution-dependent prescriptions to minimize parameter adjustments across different resolution runs. One such choice is the thermal energy floor of the unresolved ISM. \gadget provides a constant floor via the \texttt{MinEgySpec} parameter, which is resolution-independent and must be retuned by hand at each resolution level. We instead replace this specific numerical floor with a resolution-adaptive Jeans pressure floor \citep{bate97,schaye08}, applied at every cooling step alongside the \citet{springel03} effective equation of state, setting $\lambda_J=h$ in the standard Jeans criterion \citep{jeans28} and $c_s^2=\gamma(\gamma-1)u$ for an ideal gas, then solving for the minimum specific internal energy:

\begin{equation}
    u_\mathrm{Jeans} = \frac{G \rho h^2}{\pi \gamma (\gamma - 1)},
    \label{eq:jeans_floor}
\end{equation}

which naturally keeps the Jeans length larger than the smoothing length $\lambda_J \geq h$ and prevents artificial fragmentation \citep{bate97} of the cold ISM that would otherwise drive runaway star formation and gas (and dust) expulsion into the CGM and beyond at high resolution. Because $u_\mathrm{Jeans} \propto \rho h^2$, the floor scales automatically with local density and smoothing length, remaining self-consistent across the resolutions we run here ($512^3$--$2048^3$) and potentially higher without adjustment.

We run 11 versions of the model, adding one additional process with
each and labeling them S0 to S10 (see Table~\ref{tab:sim_grid}), making it possible to isolate the effect of each individual dust process relative to the previous rung. We run the complete grid at $1024^3$ resolution (gas particles of $\sim1.6\times10^6\,M_\odot$, star particles of $\sim2.3\times10^6\,M_\odot$) and a single complete (S10) run at $2048^3$ resolution, an $8\times$ reduction in particle mass for both species ($\sim2.0\times10^5\,M_\odot$ gas, $\sim2.9\times10^5\,M_\odot$ stars). Dust superparticles are spawned at creation with a mass of $\sim434\,M_\odot$ from AGB stars and $\sim1300\,M_\odot$ from core-collapse supernovae at $1024^3$, using yield fractions of $f_\mathrm{AGB}=0.4$ and $f_\mathrm{SNII}=0.2$ with 12 and 8 superparticles spawned per event, respectively. At $2048^3$, spawn counts are doubled to 24 and 16 per event to keep the dust mass per superparticle scaling roughly in step with the gas and star mass resolution scaling. This combined with the $8\times$ reduction in particle mass yields a further $16\times$ reduction in dust superparticle mass, to $\sim27\,M_\odot$ (AGB) and $\sim81\,M_\odot$ (SNII).

\subsubsection{Halo 569}
\label{sec:halo569}

We selected a single target halo, hereafter Halo 569, from the low-resolution dark matter simulation. Halo 569 was chosen because it remained well separated from the simulation boundaries, avoided nearby massive systems, and evolved into a Milky Way-like galaxy by $z=0$. Throughout this work, all radial measurements are centered on the Halo 569 friends-of-friends (FOF) halo identified in the group catalogs.

Halo 569 is followed through cosmic time by matching the $z=0$ halo to its nearest progenitor in successive snapshots. Halo properties are measured using the spherical-overdensity radius $R_{200}$, defined as the radius enclosing a mean density of $\rho_{\rm crit,200}$. Whenever possible, $R_{200}$ and $M_{200}$ are recomputed directly from the particle distribution. In a small number of snapshots, the spherical-overdensity reconstruction does not converge and the precomputed halo catalog values (\texttt{Group\_R\_Crit200} and \texttt{Group\_M\_Crit200}) are adopted instead. A few of these snapshots, concentrated around periods of merger activity at $z\sim0.65$ and $z\sim4$, show discontinuous jumps in $M_{200}$ and $M_\star$ that do not reflect real physical mass loss or gain, but rather temporary ambiguity in halo identification during this period. We note these explicitly here since they are visible as erratic structure in Figure~\ref{fig:evolution_comparison} and should not be interpreted as intrinsic physical evolution.

Halo 569 grows steadily from a low-mass progenitor at $z\gtrsim10$
into a mature galaxy by the present day. At $z\approx10$, the main progenitor has $M_{200}\approx5\times10^9\,M_\odot$ and $R_{200}\approx5$\,pkpc; by $z\approx6$ it has grown to
$M_{200}\approx8\times10^{10}\,M_\odot$, already a significant
fraction of its present-day mass. At $z=0$, the halo has $R_{200}=113.3$ pkpc and $M_{200}=1.53\times10^{11}\mathrm{M_\odot}$. The corresponding galaxy contains $\log(M_\star/\mathrm{M_\odot})=9.70$ and $\log(M_{\rm dust}/\mathrm{M_\odot})=7.29$ in the fiducial S10 $1024^3$ simulation, landing within the stellar-to-halo mass region predicted by \citealt{behroozi13} (their Figure 9a).

\subsubsection{Spatial Hashes}
\label{sec:spatial_hash}

Finding neighboring gas, star, or dust particles is needed at each timestep so we create a snapshot of particle locations at occasional intervals which partitions the high-resolution zoom region into a 3D grid. This makes it possible to search in nearby cells rather than the full simulation, saving considerable computation.


Three independent spatial hashes are maintained (gas, stars, and dust) as the simulation evolves because processes involving feedback, dust-gas coupling, radiation pressure, and grain-grain collision physics all require neighbor searches within different particle groups. The size of the cells is adaptive based on the number of particles and the desired radius of the search.

\subsubsection{Adaptive Courant Factor}

The Courant factor sets the simulation timestep as a fraction of the time it takes a sound wave to cross a gas particle's smoothing length, and so directly controls the tradeoff between numerical stability and computational speed. With \CG's addition of computationally expensive independent dust particles, a single fixed Courant value struggles to serve the whole simulation well. Early sparser timesteps readily allow a larger Courant factor, but as star and dust particle counts grow, particularly by z=2 where numerical instabilities become common when timesteps are allowed to grow too large, a smaller and safer value is required. We address this by stepping the Courant factor down at three fixed redshifts (z=20, 10, and 2), yielding four discrete regimes of increasing caution as the simulation's complexity grows (see chosen values in Table \ref{tab:new_gadget4_params}). This adaptive scheme preserves stability at low redshift without incurring the performance cost of running the full simulation at a single conservative value.


\subsubsection{Subgrid Clumping}
\label{sec:clumping}

Spatially resolving cold molecular clouds is often untenable in large scale simulations. Subgrid models understimate grain growth unless they incorporate a special treatment for unresolved cold high density regions. The term "clumping" inadvertently suggests a physical process but it is only a numerical treatment to avoid underestimating the density of the gas in molecular clouds and in turn underestimate dust activity such as grain growth, coagulation, and shattering that are directly dependent on the surrounding density. A clumping factor, C, is often introduced \citep{li21, trayford26} as a subgrid correction method. It is essentially a multiplicative boost to the local hydrogen number density which the simulation does not resolve but enables the increased dust activity expected to occur in these environments. In the case of dust grain growth, for example, the boosted density will alter the \citealt{hk11} (HK11) grain growth time scale as described in Section \ref{sec:growth}.

The clumping factors are assigned based on the critical hydrogen number density that must be reached for star formation to be allowed, namely \texttt{CritPhysDensity} in \gadget. The factor C increases from 1 in diffuse gas to 30 in actively star-forming gas which is broadly consistent with clumping factors representing molecular clouds in the literature \citep{mao19, trayford26}. We anticipate that at resolutions of $4096^3$ and higher, the need for this subgrid clumping disappears.

\subsection{Stellar Feedback}
\label{sec:feedback}

Stellar feedback is the return of energy and metals from stars to the surrounding system. We implement this process from two sources: 1) core-collapse supernovae, which is very energetic, and 2) AGB stars, which is much more gentle.

For each feedback event, we randomly distribute energy and metals to nearby gas neighbors, tracking which particles have already been heated in the current step to prevent multiple stars from stacking energy on the same particle. Similarly, each star particle is allowed one AGB and one SNe event in its lifetime. If energy cannot be immediately deposited because there are no neighbors or the heating limit of that timestep is reached, the energy is instead diverted into the star's energy reservoir and released on subsequent timesteps. 

We treat core-collapse supernova by activating feedback once during the first timestep a star particle's age falls within 3–40 Myr, corresponding to the main-sequence lifetimes of stars massive enough to explode. The number of supernovae per stellar mass formed, where the denominator $\int m\,\xi(m)\,dm$ is the Kroupa IMF-weighted total mass \citep{kroupa01} and all masses are units of solar mass ($M_{\odot}$), becomes:

\begin{equation}
\frac{N_{\mathrm{SNe}}}{M_*}
=
\frac{
0.5 \int_{8}^{100}\, m^{-2.3}\, dm
}{
\int_{0.1}^{0.5} m^{-0.3}\, dm
+
0.5
\int_{0.5}^{100} m^{-1.3}\, dm
}
\end{equation}

as $0.011 \times M_*/M_{\odot}$, or about 1 SNII for every 90 solar masses formed. The traditional SN energy of $10^{51}$ erg and a metal yield as 2 $M_{\odot}$ per SNe is applied to each feedback event. Energy is distributed stochastically to neighboring gas particles within 2 local SPH smoothing lengths of the star via the EAGLE-style \citep{schaye15} $\Delta T$-target scheme with $\Delta T_{target} = 10^7 K$. Each selected particle receives an energy fraction proportional to its kernel-weighted mass.

AGB stars distribute their metals and energy on a longer timescale that varies with their mass. We choose a single starting age for AGB metal enrichment at 100 Myr for all star particles. The average mass loss fraction is assumed to be 0.25 and the ejected metals from an IMF-weighted AGB source are calculated as $9.956 \times 10^{-3} M_{\odot} M_{\odot}^{-1}$ from the yield tables of \citealt{huscher25}. AGB wind energy is deposited proportional to the metal yield corresponding to slow winds of 20 km/s. For this paper, we track a single bulk metallicity representing total metal mass fraction, reserving individual element abundances from AGB yield tables for a future implementation.

\section{Dust Processes}
\label{sec:dust_processes}

We include an introductory treatment for all dust physics, including drag, astration, shattering, coagulation, sputtering, grain growth, shock destruction, and radiation pressure. The notable point is that dust is not tied to gas as a passive tracer, but instead exists as individual particles which can experience dynamics and grain size changes at different rates. This allows our simulation to let dust evolution progress from first principles and explore processes often ignored in dust simulations such as how radiation pressure's acceleration of dust relative to gas evolves or when individual grains can be definitely destroyed based on their own accumulated shattering history. Figure~\ref{fig:dust_quick_ref} summarizes the governing equations and key parameters for each process described in this section.


\begin{figure*}[!p]
\label{fig:dust_quick_ref}
\setlength{\fboxsep}{8pt}
\setlength{\fboxrule}{0.5pt}

\noindent\fbox{%
\begin{minipage}{\dimexpr\textwidth-2\fboxsep-2\fboxrule\relax}

  \begin{center}
    {\large\bfseries \textsc{CosmicGrain} Dust Physics Quick Reference}\\[3pt]
    {\small Governing equations and key details. More details and references for each process are given in Section~\ref{sec:dust_processes}.}
  \end{center}
  \rule{\linewidth}{0.5pt}
  \vspace{2pt}

  \setlength{\abovedisplayskip}{3pt}
  \setlength{\abovedisplayshortskip}{2pt}
  \setlength{\belowdisplayskip}{3pt}
  \setlength{\belowdisplayshortskip}{2pt}

  \begin{minipage}[t]{0.473\linewidth}
  {\small

  \noindent\textbf{[S0]\enspace Dust Creation}
  \[
    m_{\rm dust} = f_{\rm yield}\times m_{\rm metals}
  \]
  \underline{SNII}: $a_0{=}10$\,nm, $f_C{=}0.1$; \underline{AGB}: $a_0{=}100$\,nm,
  $f_C{=}0.6$; Dust superparticles/event set by params; ejected mass subtracted from host gas metallicity; isotropic velocity kick.

  \smallskip\noindent\rule{\linewidth}{0.4pt}\smallskip

\noindent\textbf{[S1]\enspace Grain Temperature}
\[
  T_{\rm eq} = \left[\frac{P_{\rm CMB}+P_{\rm ISRF}+P_{\rm gas}}
                          {C_{\rm emit}}\right]^{1/(4+\beta_{\rm eff})}
\]
$C_{\rm emit}$: radiative emission efficiency; 3 power sources influence dust temp; $P_{\rm CMB/ISRF}{=}C_{\rm emit}T^{4+\beta_{\rm eff}}$
at $T_{\rm CMB}{=}2.7(1{+}z)$\,K,
$T_{\rm ISRF}{=}17(1{+}z)^{0.25}$\,K;
$P_{\rm gas}$ is collisional gas--grain;
$T_{\rm dust}$ relaxes to $T_{\rm eq}$ over $\tau_{\rm cool}$.

  \smallskip\noindent\rule{\linewidth}{0.4pt}\smallskip

  \noindent\textbf{[S2]\enspace Gas--Dust Drag}
  \begin{align*}
    t_{\rm stop} &= \sqrt{\tfrac{\pi\gamma}{8}}\,
      \frac{a\rho_{\rm grain}}{\rho_{\rm gas}c_s}
      \times\bigl(1{+}\tfrac{9\pi\mathcal{M}^2}{128}\bigr)^{-1/2} \\
    \Delta v_k &= (v_{{\rm gas},k}{-}v_{{\rm dust},k})
      \bigl(1-e^{-\Delta t/t_{\rm stop}}\bigr)
  \end{align*}
  Supersonic correction for $\mathcal{M}{>}0.1$; exact exponential valid
  for any $\Delta t/t_{\rm stop}$; cadenced every 10 gravity steps.

  \smallskip\noindent\rule{\linewidth}{0.4pt}\smallskip

\noindent\textbf{[S3]\enspace Astration}
\[
  \Delta m_{\rm dust} = m_\star\times(D/G)_{\rm local},
  \quad r_{\rm search}=\min(r_{\rm cell},h_{\rm sml})
\]
$(D/G)_{\rm local}$: total dust mass within $r_{\rm search}$
divided by the host gas particle mass; neighbours weighted by $1/r$; consumed dust mass added to new star's mass and metallicity.

  \smallskip\noindent\rule{\linewidth}{0.4pt}\smallskip

  \noindent\textbf{[S4]\enspace Thermal Sputtering}
  \begin{align*}
    \tau_{\rm sp} &= 0.17\,{\rm Gyr}\,\frac{a_{-1}}{\rho_{-27}}
      \Bigl[\Bigl(\frac{T_0}{T}\Bigr)^{\!\omega}\!+1\Bigr],\;
      T_0{=}2{\times}10^6\,{\rm K},\;\omega{=}2.5
  \end{align*}
  $a \to a\,e^{-\Delta t/\tau_{\rm sp}}$; C grains erode ${\sim}1.5\times$ faster ($U_{\rm eff}{=}4$ vs.\
  6\,eV); sputtered mass returned to nearest gas cell.

  \smallskip\noindent\rule{\linewidth}{0.4pt}\smallskip

  \noindent\textbf{[S5]\enspace ISM Grain Growth}
  \begin{align*}
    \tau_{\rm acc} &= 6.3{\times}10^7\,{\rm yr}\;
      \frac{a}{0.1\,\mu{\rm m}}\cdot
      \frac{Z_\odot}{Z}\cdot
      \frac{10^3}{n_{\rm eff}}\cdot
      \Bigl(\frac{T_{\rm eff}}{50\,{\rm K}}\Bigr)^{\!-1/2}\!\frac{1}{S}\\
    a &\to a\,e^{f_{\rm mol}\Delta t/\tau_{\rm acc}}
  \end{align*}
  $S{=}0.3$, $T_{\rm eff}{=}20$\,K; $f_{\rm mol}\in\{0.05,0.2,0.5,0.8\}$
  diffuse$\to$SF; $D/Z$ ceiling $0.5/(1{+}0.15z)$; $\delta m$ subtracted
  from $Z_{\rm gas}$.

  } 
  \end{minipage}%
  \hfill
  \begin{minipage}[t]{0.473\linewidth}
  {\small

\noindent\textbf{[S6]\enspace Subgrid Clumping}
\[
  n_{\rm eff} = C\,n_H, \qquad C\in\{1,\,1.5,\,3,\,10,\,30\}
\]
$C$ assigned by local density from diffuse ($C{=}1$) to actively SF ($C{=}30$); new density ($n_{\rm eff}$) used by $\tau_{\rm acc}$, $\tau_{\rm coag}$, and chosen density which either coags (>) or shatters (<)

  \smallskip\noindent\rule{\linewidth}{0.4pt}\smallskip

  \noindent\textbf{[S7]\enspace SN Shock Destruction}
  \begin{align*}
    R(t) &= \xi\!\left(\frac{Et^2}{\rho}\right)^{\!1/5},\;
      \xi{=}1.033,\; E{=}10^{51}\,{\rm erg},\; t{=}0.3\,{\rm Myr} \\
    f_{\rm vol} &= \bigl(R_{\rm phys}/R_{\rm search}\bigr)^3
  \end{align*}
  $f_{\rm vol}$ makes resolution independent; $v_{\rm shock}\propto R_{\rm phys}^{-3/2}$ always from $R_{\rm phys}$; Shock efficiency $\varepsilon(v,f_C)$ blended by composition (see Table \ref{tab:shock_efficiencies}).

  \smallskip\noindent\rule{\linewidth}{0.4pt}\smallskip

\noindent\textbf{[S8]\enspace Coagulation}
\begin{align*}
  \tau_{\rm coag} &\approx 10^7\,{\rm yr}\,
    \frac{100}{n_{\rm eff}}\,\frac{0.1\,\mu{\rm m}}{a} \\
  a &\to a\,(1 - e^{-\Delta t/\tau_{\rm coag}})^{1/3}
\end{align*}
  Active where $n_{\rm eff}{>}n_{\rm thresh}$ and $T{<}3000$\,K;
  $a_{\rm max}{=}200$\,nm; superparticle mass conserved.

  \smallskip\noindent\rule{\linewidth}{0.4pt}\smallskip

  \noindent\textbf{[S9]\enspace Shattering}
  \begin{align*}
    v_{\rm shatter} &= (1+f_C)\,{\rm km\,s}^{-1} \\
    P &= 1-e^{-\Delta t/\tau_{\rm shat}};\qquad a\to a/3
  \end{align*}
  $a/3$ is MRN mass-weighted mean fragment; active in diffuse gas ($n_{\rm eff}{<}n_{\rm thresh}$); fragments below $a_{\rm min}$ destroyed, mass returned to gas.

  \smallskip\noindent\rule{\linewidth}{0.4pt}\smallskip

  \noindent\textbf{[S10]\enspace Radiation Pressure}
  \begin{align*}
    Q_{\rm pr} &= \frac{a}{a+a_0}(0.8+0.4f_C),\quad a_0{=}35\,{\rm nm}\\
    F_{\rm rad} &=(L/4\pi r^2c)\,Q_{\rm pr}\,\pi a^2\\
    f_{\rm couple} &=
      \frac{(1-e^{-\Delta t/t_{\rm stop}})\,t_{\rm stop}}{\Delta t}
  \end{align*}
Only from stars ${<}20$\,Myr; $Q_{\rm pr}$: radiation pressure efficiency; $f_{\rm couple}\in[0,1]$ corrects for terminal velocity effects when $t_{\rm stop}\ll\Delta t$; $\varepsilon_{\rm rad}{=}0.05$ calibrates for optical depth and extended stellar populations.
  } 
  \end{minipage}

\end{minipage}%
}

\caption{Quick reference of CosmicGrain dust physics. Key references: \citet{todini01}, \citet{nozawa03}, \citet{ferrarotti06}
(creation); \citet{draine84}, \citet{mathis83}, \citet{burke83}
(temperature); \citet{mckinnon18} (drag); \citet{mckinnon17},
\citet{tielens94} (sputtering); \citet{hk11}, \citet{aoyama17} (growth);
\citet{trayford26}, \citet{choban22} (clumping); \citet{sedov59},
\citet{mckee77}, \citet{bocchio14} (shocks); \citet{jones94} (shattering); \citet{draine07}, \citet{menard10} (radiation pressure).}
\end{figure*}

\subsection{Creation}

When a feedback event occurs, a fraction of the total metal mass ejected (parameters \texttt{DustYieldSNII} and \texttt{DustYieldAGB}) is locked into newly spawned dust superparticles, with the nearest gas particle's metallicity reduced accordingly. This mass is divided equally among a specific number of particles per event (parameters \texttt{DustParticlesPerSNII} and \texttt{DustParticlesPerAGB}), each placed at a random offset from the parent star within a source-dependent radial range. The extended range for SNII reflects the time delay before dust condenses in hot SN ejecta \citep{nozawa03}, while AGB grains form closer to the star in cool circumstellar envelopes.

Each grain is initialized with properties set by the feedback source. SNII grains begin at 10 nm with a carbon fraction of 0.1, reflecting carbon-poor supernovae ejecta \citep{todini_ferrara01}. AGB grains begin at 100 nm with a carbon fraction of 0.6, consistent with gentler, carbon-enriched AGB winds \citep{ferrarotti06}. The carbon fraction, which does not evolve, will impact subsequent processes including sputtering rates, radiation pressure efficiency, and shock destruction thresholds. Each grain inherits its parent star's velocity plus an outward kick (parameters \texttt{DustVelocitySNII} and \texttt{DustVelocityAGB}) in a random direction, distinct from the feedback wind speeds discussed in Section~\ref{sec:feedback}. The birth position of each grain is stored in the snapshots to analyze dust particle's movement throughout the simulation.

Though global mass and metallicity are conserved, the gas particle that contributes metals to the dust is not necessarily the same gas particle that received the metals from the feedback process. These are separate neighbor searches using the spatial hashes discussed in Section \ref{sec:spatial_hash}.

\subsection{Cooling and Heating}
\label{sec:temperature}

Since dust grain cooling times are much shorter than the simulation timesteps, we solve for the equilibrium temperature $T_\mathrm{eq}$ at each step rather than integrating $dT/dt$ directly. Since dust grain cooling times can be short relative to simulation timesteps, we compute the $T_\mathrm{eq}$ analytically and then relax the grain temperature toward it via the exact solution to Newton's law of cooling, rather than integrating $\mathrm{d}T/\mathrm{d}t$ explicitly. We adopt the modified blackbody framework of \citealt{draine07}, with emissivity $Q_\mathrm{emit} \propto a\,T^\beta$ and spectral index $\beta_\mathrm{eff} = 2 - 0.5f_C$, where $f_C$ is the carbon fraction.

Three terms contribute to the dust grain's temperature, all expressed as powers (energy per unit time). The Cosmic Microwave Background and the Interstellar Radiation Field \citep{mathis83} each provide a radiative heating power computed from their characteristic temperatures,
$T_\mathrm{CMB} = 2.7(1+z)$\,K and
$T_\mathrm{ISRF} = 17(1+z)^{0.25}$\,K respectively, via the same
modified blackbody law used for grain emission:
\begin{equation}
    P_\mathrm{CMB} = C_\mathrm{emit}\,T_\mathrm{CMB}^{4+\beta_\mathrm{eff}},
    \qquad
    P_\mathrm{ISRF} = C_\mathrm{emit}\,T_\mathrm{ISRF}^{4+\beta_\mathrm{eff}}
\end{equation}
where $C_\mathrm{emit} = 4\pi a^2 Q_\mathrm{eff}\,\sigma_\mathrm{SB}/T_\mathrm{ref}^{\beta_\mathrm{eff}}$ is the emission coefficient. The third contribution,
$P_\mathrm{gas}$, is the collisional power exchanged between the grain and the surrounding gas, which heats or cools the grain depending on the sign of $(T_\mathrm{gas} - T_\mathrm{dust})$:
\begin{equation}
    P_\mathrm{gas} = f_\mathrm{eff}\;2\,n_H\,k_B\,
    (T_\mathrm{gas} - T_\mathrm{dust})\;\alpha_T\,\bar{v}_\mathrm{th}\,\pi a^2
\end{equation}

where $\alpha_T = 0.3$ is the thermal accommodation coefficient \citep{burke83} and $f_\mathrm{eff} \in [1.0, 2.5]$ is a piecewise function which corrects for grain charge and the increasing role of electrons at high temperatures with breakpoints at $10^3$K and $10^4$K, ranging from 1.0 (cold neutral gas) to 2.5 (hot ionized gas), following the three-regime prescription of \citealt{hollenbach79} (see their section IIIe). Setting emission equal to total heating gives an equilibrium temperature:

\begin{equation}
    T_\mathrm{eq} \equiv \left[
        \frac{P_\mathrm{CMB} + P_\mathrm{ISRF} + P_\mathrm{gas}}
             {C_\mathrm{emit}}
    \right]^{1/(4+\beta_\mathrm{eff})}
    \label{eq:Teq}
\end{equation}

where $C_\mathrm{emit} = 4\pi a^2 Q_\mathrm{eff} \sigma_\mathrm{SB} / T_\mathrm{ref}^{\beta_\mathrm{eff}}$ is the emission coefficient. The grain temperature is relaxed toward $T_\mathrm{eq}$ via the exact solution to Newton's law of cooling with constant coefficients:

\begin{equation}
T_\mathrm{dust}^{n+1} = T_\mathrm{dust}^n + \left(1 - e^{-\Delta t/\tau_\mathrm{cool}}\right)\left(T_\mathrm{eq} - T_\mathrm{dust}^n\right)
\end{equation}

where $T_\mathrm{eq}$ and $\tau_\mathrm{cool}$ are treated as fixed over the timestep and the cooling timescale is approximated as the ratio of the grain thermal energy to its radiative power:

\begin{equation}
\tau_\mathrm{cool} = \frac{C_\mathrm{grain} T_\mathrm{dust}}{P_\mathrm{emit}}, 
\end{equation}

following standard treatments of dust thermal balance \citep{draine85, guhathakurta89}, bounded below by $T_\mathrm{CMB}$ and above by the composition-dependent sublimation temperature $T_{sub} = 1500+(500*f_{carb})$. Thermal coupling is implemented as a two-way exchange, with dust-gas collisions updating both the dust temperature and the internal energy of the nearest gas particle.

\subsection{Drag Force}
\label{sec:drag}

Dust grains experience drag and radiative forces that dynamically separate them from gas \citep{squire17}. Without drag, for example, most dust will continue to travel from its supernovae or AGB source and unrealistically leave the halo entirely in a few million years. We adopt the collision-based approach from \cite{epstein24} and employ a stopping timescale as shown by \cite{mckinnon18}:

\begin{equation}
\label{eq:tau_drag}
    \tau_{drag} = \frac{ \sqrt{\pi \gamma} a \rho_{gr} }{ 2 \sqrt{2} \rho_{gas} c_s}
\end{equation}

where $\rho_{gr}$ is the assumed internal density of a dust grain of 2.4 $g/cm^{3}$. We also incorporate a supersonic correction factor from \cite{mckinnon18} (their equation 9).

The drag force is implemented one-way only, as the momentum kick on the gas from a single dust particle is negligible \citep{mckinnon18}. That is, the velocity of the dust is influenced by nearby gas, but not vice versa. 

\subsection{Astration}

Some dust grains are incorporated into newly formed stars and the metals locked in the dust get recycled in the next stellar generation. Destruction of dust by astration has been shown to be lower by up to a factor of 4 compared to destruction by supernovae shocks \citep{dubois24}, but still represents a non-negligible destruction channel. 

When a new star of mass $M_\star$ forms, we search for nearby dust particles using the dust spatial hash within a search radius equal to the effective gas cell radius $r_{\rm cell} = (3M_{\rm gas}/4\pi\rho)^{1/3}$, capped at the SPH smoothing length $h$. The total dust mass consumed follows the first term of Equation 5 from \citet{dwek98}:
\begin{equation}
    M_{\rm dust,consumed} = M_\star \times \left(\frac{M_{\rm dust}}{M_{\rm gas}}\right)_{\rm local}
\end{equation}

The mass consumed is proportional to the local dust-to-gas ratio, distributed across neighbors weighted by inverse distance. Any dust particles reduced below a minimum mass threshold are destroyed, with their metal content returned to the nearest gas particle to conserve the total metal budget. The search radius scales with gas particle mass and density, making the treatment resolution-independent.

\subsection{Sputtering}
\label{sec:sputtering}

Around sufficiently hot gas, dust particles are eroded by energetic protons and alpha particles. We adopt the gas temperature of 100,000 K as the threshold above which dust begins to experience thermal erosion and 1,000,000 K as the gas temperature which completely destroys any size dust grain. This treatment was established by \cite{draine79} and expanded by \cite{tielens94} and \cite{tsai95}, with an associated power law timescale presented by \cite{mckinnon17} that depends on gas density, temperature, and grain size:

\begin{equation}
\label{eq:tau_sputter}
\tau_{sputter} = (0.17 \text{ Gyr}) \left(\frac{a}{\rho}\right) \left[\left(\frac{T_0}{T}\right)^{\omega} + 1\right]
\end{equation}

where a is the grain radius, $\rho$ is the local gas density, $T_0 = 2\times10^6\,\text{K}$ is the characteristic sputtering temperature, and $\omega$ = 2.5. This timescale governs continuous grain erosion with the radii evolving as:
\begin{equation}
\frac{da}{dt} = -\frac{a}{\tau_{sputter}}
\end{equation}

This formulation has two important improvements over a temperature-based power law timescale. First, the dependence on grain size allows grains to erode at a constant rate in radius regardless of their current size, consistent with the original \cite{draine79} calculations. Second, the density dependence allows grains in denser environments to be destroyed more rapidly, which is relevant across the wide range of gas densities encountered in zoom simulations of galaxy formation.

\subsection{Grain Growth}
\label{sec:growth}

Following \cite{hk11}, we adopt a grain–growth framework with a unique constant ($C_{sp}$) depending on whether the grain is carbonaceous ($5.59\times10^7$ yr) or silicate ($6.30\times10^7$ yr). Because we do not track individual elements in the gas particles, we treat dust composition approximately where grain growth proceeds from a generic metal reservoir rather than abundances of specific elements. Therefore, the dust particles composition is decided by its source; silicate grains from supernovae and carbon grains from AGB stars. The HK11 growth timescale (their equations 23 and 24) is:

\begin{equation}
\begin{split}
\tau_{\rm acc} = C_{\rm sp} 
&\left(\frac{a}{0.1\,\mu{\rm m}}\right)
\left(\frac{n_{\rm H}}{10^3\,{\rm cm}^{-3}}\right)^{-1}
\left(\frac{T}{50\,{\rm K}}\right)^{-1/2} \\
&\times\left(\frac{Z}{Z_\odot}\right)^{-1}
\left(\frac{S}{0.3}\right)^{-1}
\end{split}
\end{equation}

where $C_{\rm sp} = 6.30 \times 10^7$\,yr for silicate and $5.59 \times 10^7$\,yr for carbonaceous grains, $a$ is the grain radius, $n_{\rm H}$ is the hydrogen number density, 
$T$ is the gas temperature evaluated at the fixed temperature of $20$\,K representing cold molecular cloud conditions we are unable to resolve, $Z/Z_\odot$ is the gas metallicity relative to solar ($Z_\odot = 0.02$), and $S = 0.3$ is the sticking coefficient \citep{hk11}.

The grain radius is updated each timestep as
\begin{equation}
\label{eq:growth_radius}
    a^{n+1} = a^{n} \exp\!\left(\frac{f_{\rm mol}\,\Delta t}{\tau_{\rm acc}}\right),
\end{equation}
where $f_{\rm mol}$ is the subgrid molecular fraction of the gas and $\Delta t$ is the simulation timestep. Following \citet{aoyama17}, we adopt $f_{\rm mol}$ to improve the HK11 treatment by accounting for the fraction of mass in each gas particle that actually could accrete mass. Rather than using a fixed value as in \citet{aoyama17}, we make $f_{\rm mol}$ density-dependent: star-forming cells ($\dot{M}_\star > 0$) are assigned $f_{\rm mol} = 0.8$, reflecting their predominantly molecular composition, while diffuse cells receive $f_{\rm mol} = 0.05$, with intermediate values at moderate densities. The subgrid clumping factor, discussed in Section \ref{sec:clumping}, is what creates the needed molecular densities and enables any grain growth to occur.

The mass increase as a result of the new radius from Equation \ref{eq:growth_radius} is transferred directly from the gas particle's total mass to the dust particle. The mass of the gas particle is reduced by exactly the dust particle's increased mass, ensuring baryonic mass conservation, and its metallicity is recomputed from the remaining metal content:
\begin{equation}
    Z_{\rm new} = (M_{\rm gas} Z_{\rm gas} - dm)/M_{\rm gas,new}
\end{equation}

We implement a redshift-dependent cap on the acceptable dust to metallicity ratio to manage excessive growth in low metallicity gas. A density threshold is also implemented to prevent growth in gas that is too diffuse (regions such as the CGM). We apply our own calibration parameter as a multiplicative factor on the growth timescale to tune the growth activity to reproduce reasonable Dust-to-Gas ratios (see Table \ref{tab:new_gadget4_params}).

\subsection{Destruction by Shocks}
\label{shocks}

Supernova explosions both distribute dust outward and destroy new and existing dust by shocking the surrounding gas. We employ the classic Sedov-Taylor point explosion treatment \citep{taylor50} which models a fixed amount of energy expanding into a uniform medium of mass density $\rho$. The radius and the velocity of the shock as a function of time are:

\begin{equation}
R(t) = \xi \left( \frac{E\, t^{2}}{\rho_0} \right)^{1/5}
\end{equation}

We treat each supernova event as instantaneous in the simulation, however, so it is useful to calculate the blast velocity in a time-independent form as a function of blast radius and gas density by substituting $t = \xi^{-5/2}\left(\frac{R^{5}\rho_0}{E}\right)^{1/2}$ and eliminating $t$ giving a velocity that depends only on the instantaneous blast radius and local gas density:
\begin{equation}
v_{\rm s}(R) = \frac{2}{5}\,\xi^{5/2} \sqrt{\frac{E}{\rho_0}}\, R^{-3/2},
\end{equation}

where $\xi = 1.033$ is the Sedov self-similarity constant \citep{sedov59}.

The shock velocity is the primary variable for the efficiency of dust destruction and grains that survive are partially eroded. Much work has been done in this area including material science in the laboratory (Jones et al 1994, 1996), theoretical modeling \citep{bocchio14}, and supernova remnant observations \citep{slavin15}. We adopt the data for the erosion efficiency as a function of local shock velocity from \citep{bocchio14}, who present separate values for carbon and silicon-based grains (their Table 6). In our model, we interpolate linearly between the tabulated shock velocities (see Table~\ref{tab:shock_efficiencies}). Mixed grains blend the two efficiencies by carbon fraction $f_C$:
\begin{equation}
    \varepsilon(v, f_C) = f_C\,\varepsilon_{\rm carb}(v) 
                                   + (1 - f_C)\,\varepsilon_{\rm sil}(v).
\end{equation}
The silicate efficiency is held constant at 0.67 above 200 km\,s$^{-1}$, the upper limit of the \citet{bocchio14} table; carbonaceous grains saturate at 1.0 above 175 km\,s$^{-1}$.

\begin{table}
\centering
\caption{Shock destruction efficiency $\varepsilon$ as a function of shock velocity for carbonaceous and silicate grains from \citet{bocchio14}. Values between tabulated velocities are linearly interpolated in the simulation.}
\label{tab:shock_efficiencies}
\begin{tabular}{ccc}
\hline\hline
$V_{\rm shock}$ (km\,s$^{-1}$) & Carbonaceous & Silicate \\
\hline
 50  & 0.77 & 0.02 \\
 75  & 0.83 & 0.12 \\
100  & 0.91 & 0.29 \\
125  & 0.96 & 0.46 \\
150  & 0.99 & 0.53 \\
175  & 1.00 & 0.67 \\
200  & 1.00 & 0.67 \\
\hline
\end{tabular}
\end{table}

The shock radius is evaluated at a characteristic time of t=0.3 Myr, giving a representative Sedov-phase radius of $\sim$30--80 pc at typical ISM densities. To prevent unphysically large shock radii, the ambient density used in the Sedov-Taylor calculation is capped at $n_{\rm amb} = 0.1\,{\rm cm}^{-3}$ (see Table \ref{tab:new_gadget4_params}), which corresponds to a minimum shock velocity of $\sim96$ km\,s$^{-1}$ which is well above the 50 km\,s$^{-1}$ destruction threshold.

At all of our target resolutions, the physical shock radius ($\sim0.03-0.05 kpc$) is smaller than the dust hash cell size. We therefore use a larger search radius for finding nearby dust particles and adjust the volume accordingly for those that will actually feel the shock with a basic correction factor:

\begin{equation}
    f_{\rm vol} = (R_{\rm phys}/R_{\rm search})^3
\end{equation}

This makes the destruction rate resolution-independent: at lower resolution the $f_{\rm vol}$ is larger but fewer grains are found; at higher resolution the $f_{\rm vol}$ is smaller but more grains are found, and the mass to be destroyed converges to the same physical amount.

With the search volume, physical volume, and destruction efficiencies in hand, we sum the total dust mass within the search volume. This dust mass to destroy is distributed across all found grains, and any below a threshold radius (1nm) are completely destroyed.


\subsection{Coagulation}
\label{sec:coagulation}

Grain coagulation occurs when dust particles collide and stick together in dense and cold environments, growing grain sizes without depleting the gas-phase metals. We implement coagulation following the timescale formulation of \citet{hk11}, where the coagulation timescale scales as:

\begin{equation}
    \tau_{\rm coag} \sim 10\,{\rm Myr}
    \times \left(\frac{100\,{\rm cm}^{-3}}{n_{\rm eff}}\right)
    \times \left(\frac{0.1\,\mu{\rm m}}{a}\right),
    \label{eq:tau_coag}
\end{equation}

where $n_{\rm eff} = n_H \times C$ is the density enhanced by the clumping factor (Section~\ref{sec:clumping}), $a$ is the grain radius, and the timescale is further scaled by a calibration parameter $\mathcal{C}_{\rm coag}$. The inverse dependence on gas density reflects that in denser gas, dust grains encounter each other more frequently, reducing the coagulation time. Smaller grains coagulate faster due to their higher surface area per unit mass \citep{zhukovska08}.

Coagulation is restricted to cold ($T < 3000$\,K), dense ($n_{\rm eff} > n_{\rm thresh}$) environments targeting molecular cloud conditions \citep{asano13}. While physically molecular clouds have temperatures of $\sim$10--50\,K, the 3000K ceiling accounts for SPH kernel smoothing at finite resolution which averages over unresolved cold clumps; the coagulation timescale (Equation~\ref{eq:tau_coag}) encodes
the correct density dependence regardless. The density threshold is scaled with resolution as

\begin{equation}
    n_{\rm thresh} \approx 150 \times \rho_{\rm crit},
    \label{eq:nthresh}
\end{equation}

where $\rho_{\rm crit}$ is the star formation density threshold, ensuring consistent activation across our $512^3$--$2048^3$ runs ($n_{\rm thresh} = 3$, $20$, and $30$\,cm$^{-3}$ respectively). This range is consistent with the literature, which broadly adopts coagulation thresholds of $n_H \sim 10$--$100$\,cm$^{-3}$ \citep{hk11, zhukovska08, asano13, yan04}.

The grain radius update is mass-conserving: the superparticle represents a population of grains of radius $a$ before coagulation and a smaller number of larger grains of radius $a_{\rm new}$ after, with total dust mass unchanged. Grains are capped at $a_{\rm max} = 200$\,nm, consistent with the upper end of the standard size distribution from \citealt{mathis77} ("MRN"). In our model, coagulation occurs only when a grain's local "effective" density ($n_{\rm eff}$) is at least $5\times$ above the minimum star formation threshold, meaning the gas must be denser than the star formation floor even after the clumping boost.

\subsection{Shattering}
\label{sec:shattering}

Grain shattering is the complementary process to coagulation (Section~\ref{sec:coagulation}). Coagulation operates in cold dense gas ($n_{\rm eff} > n_{\rm thresh}$) where dust grains impact gently and stick together while shattering operates in warm diffuse gas ($n_{\rm eff} < n_{\rm thresh}$) where collisions have too much energy and grains instead fragment into pieces. We again employ the timescale formulation of Equation \ref{eq:tau_coag} from \citet{hk11}.

The turbulent velocity is approximated by the local sound speed, which is reasonable in the warm diffuse ISM where shattering operates. Shattering requires the turbulent velocity to exceed a composition-dependent threshold of $v_{\rm thresh} = 1 + f_C \quad {\rm km\,s}^{-1}$ \citep{jones94} giving 1 km\,s$^{-1}$ for pure silicate grains and 2 km\,s$^{-1}$ for pure carbonaceous grains. The timescale is reduced for grains well above the velocity threshold, with a factor $(1 + v_{\rm excess}/v_{\rm thresh})^{-1}$ representing the increased fragmentation rate at higher collision velocities.

Unlike coagulation, shattering is treated stochastically: in each timestep a grain either fragments or does not, with Poisson probability $P = 1 - \exp(-dt/\tau_{\rm shat})$. When fragmentation occurs, the grain radius is reduced by a factor of 3, consistent with the mean fragment size from an MRN $a^{-3.5}$ distribution \citep{jones96}. Total superparticle mass is conserved — the grain population now consists of more, smaller grains represented by the new radius. The timescale is further scaled by a calibration parameter DustShatteringCalibration (see Table \ref{tab:new_gadget4_params}).

\subsection{Radiation Pressure}
\label{sec:radiation_pressure}

One advantage of modeling explicit dust particles is that we can implement unique velocities for dust due to radiation pressure from nearby stars. Following \citet{draine84}, the radiation pressure force on a single grain is:

\begin{equation}
    F_{\rm rad} = \frac{L}{4\pi r^2 c}\, Q_{\rm pr}\, \pi a^2
\end{equation}

where $a$ is the grain radius and the flux $L/4\pi r^2 c$ assumes optically thin conditions between star and grain. Deviations from the the optically thin simplification, such as dust self-shielding in
denser regions and the fact that each star particle represents an
extended stellar population, are absorbed
into a single calibration factor $\epsilon_\mathrm{rad}$.

The force direction is computed as the unit vector from each stellar particle to the dust grain, with contributions accumulating from all young neighbors (< 20 Myr) within the search radius. We adopt a simplified approximation for the radiation pressure efficiency which provides a modest 1.5x composition dependence between pure carbon and pure silicate grains:

\begin{equation}
Q_{\rm pr} = \frac{a}{a + a_0}\,(0.8 + 0.4\,f_C),
\end{equation}

where $a_0 = \lambda/2\pi = 35\,{\rm nm}$ corresponds to the UV-dominated radiation field that contributes most to radiation pressure, and $f_C$ is the carbon fraction capturing the higher absorption efficiency of carbonaceous grains relative to silicates \citep{draine84}. This approach reproduces the qualitative behavior of their Figures 4 and 5.

As we do not explicitly model the radiation field, we adopt a simplified prescription for the stellar luminosity based on population synthesis models \citep{leitherer99}. We approximate the bolometric luminosity of each star particle using an age-dependent mass-to-light ratio, assuming a Kroupa IMF averaged over the full stellar population rather than individual O-star luminosities. Stars older than 20 Myr are excluded, as the OB stellar population responsible for UV-driven dust coupling has largely dispersed by this age.

\begin{equation}
    \frac{L}{M} = \begin{cases}
        200 \, L_\odot \, M_\odot^{-1} & t_* < 3 \, \text{Myr} \\
        100 \, L_\odot \, M_\odot^{-1} & 3 \leq t_* < 10 \, \text{Myr} \\
        20  \, L_\odot \, M_\odot^{-1} & 10 \leq t_* < 20 \, \text{Myr} \\
        0                              & t_* \geq 20 \, \text{Myr}
    \end{cases}
\end{equation}

The resulting acceleration is scaled by the factor
$\epsilon_\mathrm{rad}$, calibrated to $\epsilon_\mathrm{rad} = 0.05$
such that the predicted dust surface density profile follows the
observed $\Sigma_\mathrm{dust} \propto r^{-0.8}$ slope beyond 30 kpc
reported by \citet{menard10}.

In order to keep the radiation pressure process scalable with resolution, we set the search radius for nearby dust particles to scale with gas smoothing length, including a floor of $2 \times$ the star hash cell size.


\begin{deluxetable*}{llcccccccccc}
\tablecaption{%
  Dust observables for the CosmicGrain simulation ladder at $1024^3$ resolution.
  \label{tab:dust_sim_ladder_1024}}
\setlength{\tabcolsep}{4pt}
\tablewidth{0pt}
\tablehead{
  \colhead{Sim} &
  \colhead{Description} &
  \colhead{$M_\star/M_{\rm halo}$} &
  \colhead{$M_{\rm dust}/M_\star$} &
  \colhead{$f_{\rm surv}$} &
  \colhead{$\langle a\rangle_{\rm ISM}$} &
  \colhead{$\langle a\rangle$} &
  \colhead{$D/G|_{\rm ISM}$} &
  \colhead{$D/G$} &
  \colhead{$D/Z|_{\rm ISM}$} &
  \colhead{$D/Z$} &
  \colhead{$\bar{f}_{\rm C}$}
}
\startdata
  S0 & Creation only & 0.030 & $1.9\times10^{-3}$ & 1.00 & 32.4 & 40.2 & $2.7\times10^{-4}$ & $8.1\times10^{-4}$ & 0.02 & 0.11 & 0.27 \\
  S1 & $+$ Dust cooling & 0.028 & $2.0\times10^{-3}$ & 1.00 & 39.5 & 40.0 & $2.5\times10^{-4}$ & $8.5\times10^{-4}$ & 0.01 & 0.11 & 0.27 \\
  S2 & $+$ Gas--dust drag & 0.031 & $6.3\times10^{-3}$ & 1.00 & 15.2 & 35.2 & $6.7\times10^{-2}$ & $2.8\times10^{-3}$ & 0.84 & 0.29 & 0.24 \\
  S3 & $+$ Astration & 0.029 & $4.5\times10^{-3}$ & 0.86 & 13.5 & 25.4 & $3.4\times10^{-2}$ & $1.6\times10^{-3}$ & 0.70 & 0.22 & 0.19 \\
  S4 & $+$ Thermal sputtering & 0.032 & $1.2\times10^{-3}$ & 0.36 & 40.3 & 52.3 & $2.9\times10^{-3}$ & $5.0\times10^{-4}$ & 0.31 & 0.06 & 0.36 \\
  S5 & $+$ ISM grain growth & 0.028 & $1.5\times10^{-3}$ & 0.39 & 43.1 & 53.0 & $3.1\times10^{-3}$ & $5.6\times10^{-4}$ & 0.17 & 0.07 & 0.36 \\
  S6 & $+$ Subgrid clumping & 0.032 & $5.9\times10^{-3}$ & 0.29 & 16.6 & 19.9 & $7.4\times10^{-2}$ & $2.5\times10^{-3}$ & 0.81 & 0.29 & 0.13 \\
  S7 & $+$ SN shock destr. & 0.032 & $5.7\times10^{-3}$ & 0.28 & 16.5 & 20.3 & $7.9\times10^{-2}$ & $2.7\times10^{-3}$ & 0.75 & 0.28 & 0.13 \\
  S8 & $+$ Coagulation & 0.032 & $4.8\times10^{-3}$ & 0.30 & 16.6 & 20.0 & $4.8\times10^{-2}$ & $2.1\times10^{-3}$ & 0.66 & 0.22 & 0.13 \\
  S9 & $+$ Shattering & 0.033 & $3.3\times10^{-3}$ & 0.26 & 13.2 & 17.3 & $3.3\times10^{-2}$ & $1.6\times10^{-3}$ & 0.56 & 0.16 & 0.14 \\
  S10 & $+$ Radiation pressure & 0.028 & $2.9\times10^{-3}$ & 0.32 & 13.9 & 34.8 & $1.4\times10^{-2}$ & $1.1\times10^{-3}$ & 0.41 & 0.13 & 0.24 \\
    \hline
    S10 & $512^3$ & 0.023 & $3.1\times10^{-3}$ & 0.40 & 46.6 & 39.2 & $1.5\times10^{-2}$ & $1.2\times10^{-3}$ & 0.30 & 0.23 & 0.26 \\
    S10 & $1024^3$ & 0.028 & $2.9\times10^{-3}$ & 0.32 & 13.9 & 34.8 & $1.4\times10^{-2}$ & $1.1\times10^{-3}$ & 0.41 & 0.13 & 0.24 \\
    S10 & $2048^3$ & 0.032 & $3.1\times10^{-3}$ & 0.86 & 33.3 & 54.1 & $6.1\times10^{-2}$ & $1.2\times10^{-2}$ & 0.33 & 0.35 & 0.31 \\
  \hline
  MW & Observed & $\sim0.04$ & $5\times10^{-4}$ & \nodata & \nodata & \nodata & $7\times10^{-3}$ & $7\times10^{-3}$ & 0.45 & \nodata & \nodata \\
  \hline
\enddata
\tablecomments{%
Each rung adds one physics process to the previous configuration. All runs use identical initial conditions (50 Mpc comoving box with smaller high-resolution zoom region) and follow the evolution of Halo 569. Halo properties are measured using the spherical-overdensity radius $R_{200}$ and mass $M_{200}$ returned by the halo tracking procedure described in Section \ref{sec:halo569}. At z=0, the halo masses of the ladder runs span $M_{200}\approx1.5-1.8\times10^{11}\,h^{-1}\,M_\odot$ with $R_{200}\approx113-120$ pkpc. $M_\star/M_{\rm halo}$: stellar-to-halo mass ratio within $R_{200}$. $M_{\rm dust}/M_\star$: dust-to-stellar mass ratio within $R_{200}$. $f_{\rm surv}$: surviving dust fraction $N_{\rm alive}/(N_{\rm alive}+N_{\rm destroyed,cumul})$. $\langle a\rangle_{\rm ISM}$: mass-weighted mean grain radius for dust in the ISM-like radius ($r<20\,{\rm pkpc}$). $\langle a\rangle_{200}$: same, for all dust within $R_{200}$. $D/G|_{\rm ISM}$, $D/Z|_{\rm ISM}$: dust-to-gas and dust-to-metals ratios restricted to the ISM aperture above. $D/G$, $D/Z$: same ratios using total gas and metals within $R_{200}$.  $\bar{f}_{\rm C}$: mass-weighted mean carbon fraction within $R_{200}$. Milky Way reference values are shown for comparison: $M_\star/M_{\rm halo}$ from \citet{blandhawthorn16}; $D/G$ and $M_{\rm dust}/M_\star$ from \citet{draine07}; $D/Z|_{\rm ISM}$ from \citet{jenkins09} which is derived from local ISM depletion measurements and is most directly comparable to the inner 20 pkpc.
}
\end{deluxetable*}

\section{Results}
\label{sec:results}

The $1024^3$ runs require $\approx100$ core hours and the $2048^3$
runs require $\approx4000$ core hours to complete. Runtime across the
ladder varies by $\lesssim10\%$, with the latest rungs (S9, S10) taking
modestly longer despite having roughly half the final dust particle
count of the earliest rungs (destruction channels becoming active),
indicating that the additional grain-grain physics (shattering,
coagulation, radiation pressure), more than the raw particle count, drives
the runtime difference.

We monitor global baryon mass conservation properties throughout each run by summing the masses of all particle species in every output snapshot and the combined baryonic mass (gas, stars, and dust) is conserved to better than $0.001\%$ ($\sim 2.5 \times 10^7 M_{\odot}$) over the full simulation. The residual discrepancy is likely due to a combination of numerical effects and bookkeeping within the subgrid dust model and we consider to be negligibly small for the scientific analysis presented here.

\subsection{Dust Ladder}

In Table \ref{tab:dust_sim_ladder_1024}, we present numerical results at z=0 for the full dust physics ladder focused on one Milky Way-size halo, Halo 569, with one additional dust process added on each rung. This is a useful approach to both uncover bugs and to compare individual components in this complex system. While the resulting observables should not be treated as absolute calibrated values, they do show the relative changes due to each process of dust physics. We calculate the dust-to-gas (D/G) and dust-to-metal (D/Z) ratios both within the ISM and within R200, with the ISM for our halo extending to 20 pkpc where the stellar surface density levels off.
\vspace{1mm}

\noindent\textbf{S0}: The baseline run establishes a stellar-to-halo mass ratio of 0.030 and a dust-to-stellar mass ratio of $1.9 \times 10^{-3}$. With no destruction mechanisms active, the survival fraction is naturally $f_{\rm surv}=1.00$. Very little dust remains in the halo as it travels outward unimpeded.
\vspace{1mm}

\noindent\textbf{S1}: Adding dust cooling produces only minor changes to the global dust population, with dust-to-stellar and dust-to-metals ratios remaining nearly unchanged from S0. The $\sim20\%$ increase in ISM grain radius is noise in a small remaining sample as little dust remains within 20 kpc.
\vspace{1mm}

\noindent\textbf{S2}: Gas--dust drag substantially increases dust retention within $R_{200}$, raising the global dust-to-stellar mass ratio by a factor of $\sim3$ and the ISM dust-to-gas ratio by more than 2 orders of magnitude. Nearly all of the metals in the ISM are in dust.
\vspace{1mm}

\noindent\textbf{S3}: The inclusion of astration produces a substantial suppression of the dust population. The survival fraction $f_{\rm surv}$ drops from 1.00 to 0.86, indicating astration destroys a meaningful fraction ($\sim14\%$) of dust superparticles. Global dust-to-gas and dust-to-metal ratios decline by 43\% and 24\% respectively relative to S2, with comparable reductions in the ISM values. The dust-to-stellar-mass ratio also decreases by $\sim30$\%, consistent with astration preferentially removing dust that would otherwise continue growing in the dense star-forming gas where new stars form.
\vspace{1mm}

\noindent\textbf{S4}: Thermal sputtering strongly suppresses the dust budget, reducing $M_{\rm dust}/M_\star$ by roughly three quarters and lowering the survival fraction to $f_{\rm surv}=0.36$. The mean grain size and carbon fraction both increase substantially, consistent with preferential destruction of smaller and less resilient grains.
\vspace{1mm}

\noindent\textbf{S5}: Grain growth produces only modest increases in dust mass and grain size. At this resolution, the resolved ISM densities are insufficient for accretion to substantially replenish the dust destroyed by sputtering.
\vspace{1mm}

\noindent\textbf{S6}: Introducing subgrid clumping significantly enhances grain growth, increasing the global dust-to-stellar mass ratio by a factor of $\sim4$ and restoring the dust-to-metals ratio to values comparable to the earlier drag-dominated models. The mean carbon fraction declines sharply as grain growth preferentially builds silicate mass.
\vspace{1mm}

\noindent\textbf{S7}: Adding supernova shock destruction produces almost no change relative to S6 in any dust observable, including the surviving dust fraction, global dust budget, and dust-to-metals ratios. This near-total insensitivity to an added destruction channel is counterintuitive and is discussed further in Section~\ref{sec:discussion}.
\vspace{1mm}

\noindent\textbf{S8}: Coagulation reduces the $R_{200}$ dust mass modestly, while the characteristic grain size remains essentially unchanged. Dust retention in the ISM declines, with both global dust-to-gas and dust-to-metals ratios decreasing relative to S7.
\vspace{1mm}

\noindent\textbf{S9}: Shattering further reduces the dust budget and lowers the surviving dust fraction to $f_{\rm surv}=0.26$. The mean grain size decreases rather than increases at this rung, suggesting shattering directly fragments grains across the surviving population rather than preferentially removing only the smallest.
\vspace{1mm}

\noindent\textbf{S10}: Radiation pressure strongly depletes dust from the central ISM, reducing $D/G_{\rm ISM}$ by more than a factor of two and $D/Z_{\rm ISM}$ by roughly one third. The halo-wide mean grain size more than doubles even as the ISM mean grain size is nearly unchanged, indicating radiation pressure selectively transports larger grains out of the ISM into the CGM rather than destroying dust overall.
\vspace{1mm}

\section{Discussion}
\label{sec:discussion}

A key structural feature of CosmicGrain, made possible by representing dust as explicit superparticles rather than a post-processed or subgrid-averaged dust field, is that dust processes like drag coupling or thermal sputtering can respond to the \emph{local} gas state at each individual grain's position, instead of to galaxy-averaged properties. Each dust superparticle's drag timescale (Eq. \ref{eq:tau_drag}) and sputtering timescale (Eq.~\ref{eq:tau_sputter}) is evaluated with its own nearest gas neighbor's density, temperature, and relative velocity, identified via the spatial hash (Section~\ref{sec:spatial_hash}). This allows the model to produce a genuinely heterogeneous outcome across the grain population.

For example, a size-dependent sputtering asymmetry arises because individual grains experience different local conditions depending on exactly where they exist relative to the densest, hottest gas, not because of a broader size-dependent prescription. Models that instead evolve dust as a passive tracer of gas, or apply a single galaxy-averaged sputtering rate, cannot capture this heterogeneity. The same advantage extends to grain growth, coagulation, and shattering, all of which depend explicitly on a grain's current size which is itself the cumulative record of all prior growth, sputtering, coagulation, and shattering events.

\subsection{Dust Physics}
\label{sec:dust_physics}

\begin{figure*}
    \includegraphics[width=\textwidth]{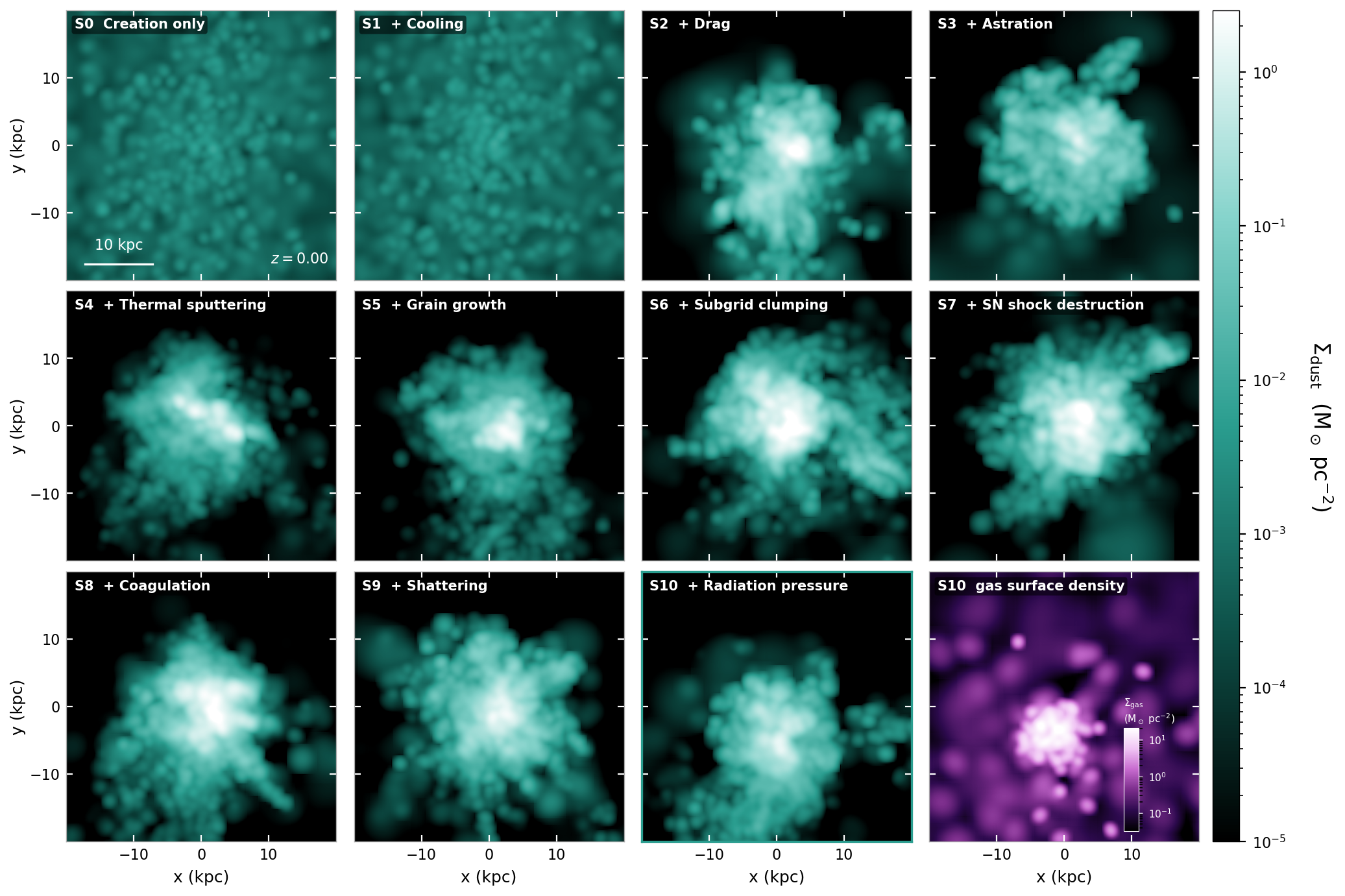}
    \caption{ 
Dust surface density in each of the 11 runs at z=0 projected along the $z$-axis to a depth of 20 kpc, centered on Halo 569, with adaptive smoothing at a 0.5 kpc scale. Panel 12 shows the S10 gas surface density with the same adaptive projection for comparison.
    }
    \label{fig:surface_densities}
\end{figure*}

\textbf{Gas-dust drag:} Of all physics processes in the ladder, drag has the largest qualitative effect on where dust resides. Figure \ref{fig:surface_densities} shows that without drag (S0--S1), essentially no dust survives in the ISM at all. Only when drag is enabled in S2 does an ISM dust population remain. Quantitatively, S2 triples the dust mass within $R_{200}$ relative to S1 and dramatically increases $D/G$ and $D/Z$, as dust grains can now couple to the surrounding gas and be retained within the halo, particularly in the ISM. These values are significantly higher than observed Milky Way values (see orange dot in Figure \ref{fig:dz_vs_zgas}), which is expected at this stage since no destruction channel has yet been introduced. The mass-weighted mean grain radius drops by $62\%$ in the ISM aperture but only $12\%$ within $R_{200}$ overall. Because larger grains have proportionally longer stopping times, they decouple from the gas more readily and continue along outward trajectories from their initial birth rather than being decelerated and retained near the disk, depleting the ISM of large grains specifically.

\textbf{Astration:} Unlike the other destruction processes introduced later in the ladder, astration removes dust mass proportionally to the local dust-to-gas ratio as stars form, rather than destroying superparticles outright. A dust grain is only fully destroyed if this removal drops its mass below the destruction floor. At S3, this floor is now crossed for a meaningful fraction of the dust population: $f_{\rm surv}$ drops from 1.00 (S2) to 0.86, with astration fully destroying 217,878 superparticles (totaling $1.36\times10^{7}\,M_\odot$) over the course of the S3 run. This is a modest but genuine destruction channel, distinct from a purely partial, ongoing mass loss. Its effect remains secondary, however, to thermal sputtering: introducing thermal sputtering at S4 drops $f_{\rm surv}$ a further 50 percentage points, to 0.36, and thermal sputtering remains the dominant destruction channel for the remainder of the ladder ($f_{\rm surv}=0.26$--$0.39$ from S4 through S10). Astration therefore contributes a real but comparatively minor share of total dust destruction in \CG, subordinate to thermal sputtering once the latter is active.

\textbf{Thermal sputtering:} The dust population experiences its strongest reducing process in sputtering, clearly visible from the red line and dot in Figures \ref{fig:dust_ratios} and \ref{fig:dz_vs_zgas}, respectively. Smaller grains are destroyed faster from sputtering: the average grain size roughly doubles in the halo and nearly triples in the ISM at S4. Building on astration's already-reduced population, this rung drives the survival fraction down substantially further, from $f_{\rm surv}=0.86$ (S3) to $0.36$ -- the largest single-rung reduction anywhere in the ladder. Carbonaceous grains have a lower sputtering binding energy ($\sim4$ eV) than silicates ($\sim6$ eV) \citep{tielens94, nozawa06}, so carbon grains should sputter faster at a given grain size, driving $f_{\rm carb}$ down. However, $\bar{f}_{\rm C}$ instead rises from $0.19$ to $0.36$, the highest value anywhere in the ladder, because smaller grains do not survive as long and silicates make up most of the smaller population despite their higher binding energy. Because the \citealt{mckinnon17} timescale is proportional to grain radius and the radius decays exponentially, smaller grains (which have shorter timescales) lose a larger fraction of their radius per unit time than larger grains do, even though sputtering removes a similar absolute thickness per unit time across all sizes.

This same destruction asymmetry shapes the surviving dust population's age distribution at $z=0$ (Figure~\ref{fig:dust_properties}). Grains rise steeply toward ages of $\sim10$--$13$\,Gyr, reflecting the long-lived AGB-carbon population, which survives longest because it is born larger, outweighing its lower binding energy. The near absence of young ($<2$\,Gyr) grains indicates that recent star formation contributes little to the surviving dust population, consistent with the declining star formation rate after cosmic noon and with newly-formed grains being destroyed by thermal sputtering before they can grow to sizes that would allow them to survive.

\textbf{Grain growth:} $D/G$ and $D/Z$ in the halo rise modestly in step with the dust mass increase ($+12\%$ and $+17\%$), consistent with grain growth transferring mass from the gas-phase metal budget onto existing grains without net creation or destruction of metals -- total metal mass is conserved, so any mass moved into dust necessarily raises the halo-wide $D/Z$ rather than lowering it. Within the smaller ISM radius, however, this conservation argument does not directly apply: $D/G_{\rm ISM}$ rises only slightly ($+7\%$) while $D/Z_{\rm ISM}$ actually declines by $45\%$, indicating that gas-phase transport into and out of the 20 pkpc ISM region outweighs the local grain-growth signal at this rung.

A sample of individual HK11 growth activity from the S5 run (3,394 logged events: 2,489 carbon, 905 silicate) shows that silicate grains grow moderately faster than carbon on a per-event basis, with growth for both species -- and particularly for silicate -- dominated by a comparatively small number of rare, high-density accretion episodes rather than steady accumulation under typical local conditions. This suggests that at this resolution higher gas densities are needed to drive efficient accretion, which is exactly the gap subgrid clumping is designed to address.

\textbf{Subgrid clumping:} Figure~\ref{fig:surface_densities} shows the
visual impact of artificially boosting gas density on dust production as the S6 panel shows a markedly denser and more extended dust distribution than S5. The HK11 growth event sample confirms the impact directly, with median $n_{\rm eff}$ rising by a factor of $9.3$ for carbon and $7.6$ for silicate with clumping enabled. Increasing the gas density in this way nearly quadruples the dust mass and increases the ISM $D/G$ and $D/Z$ ratios by factors of $24$ and $4.8$, with corresponding halo-wide increases of $4.5\times$ and $4.1\times$. Despite carbon's larger nominal density boost, silicate grains absorb most of clumping's benefit: the median silicate-to-carbon fractional growth ratio widens from $1.4\times$ without clumping to $2.9\times$ with it, consistent with silicate already sampling shorter accretion timescales at baseline and being better positioned to capitalize on any further density increase. This directly explains why the mean carbon fraction declines as grain growth preferentially builds silicate mass.

\textbf{SN shock destruction:} S7 produces a small increase in dust mass
rather than the expected decrease from an added destruction process. No grains are being destroyed across the entire run currently, and partial grain erosion sums to just $3.3\times10^{-7}\,M_\odot$,
essentially negligible. This undesirable inaccuracy is due to a spatial-resolution limitation rather than a numerical bug. The SN search radius is fixed at $3$\,kpc to keep the dust spatial hash usable, while the true Sedov-Taylor blast radius is $\sim$50\,pc, a factor of $\sim$60 smaller. To accommodate for the difference, we implement a volume-correction factor, $f_{\rm vol}\sim(R_{\rm phys}/R_{\rm search})^3$ which works out to be $\sim5.8\times10^{-6}$. The dust found near the supernovae, when multipled by this tiny factor, suppresses the destroyed mass too strongly.

We calculate that refining the spatial hash enough to directly resolve a 50\,pc blast radius across the zoom region would require a grid far beyond any computationally feasible size. The current subgrid treatment also assumes dust is uniformly distributed within the oversized search volume. As real dust is likely bunched up near sites of star formation, this assumption likely underestimates destroyed mass as well. We defer a resolution of this limitation, potentially involving a subgrid treatment separate from the shared spatial hash, to future work. The carbon fraction still declines modestly at this rung, consistent with carbon grains' lower binding energy making them more susceptible than silicates to whatever shock destruction efficiency the model does manage to apply \citep[Table 6 of][]{bocchio14}. The lack of dust destruction this mechanism would otherwise impart contributes to the result of Figure \ref{fig:stellar_vs_dust_mass}, where Halo 569 and the other satellites tend to have somewhat more dust mass than observational and model comparisons at the same stellar mass.

\textbf{Coagulation:} Across the full run, coagulation grows grains in $302{,}154$ total events, mostly within actively star-forming gas which is the model's maximum clumping tier ($C=30$). Rarer events in dense but non-star-forming gas (clumping factors of $C=10$ and $C=3$) also contribute, though the lowest clumping tiers ($C=1$-$1.5$) for diffuse gas apparently never satisfy coagulation's density and temperature gates in this run, with a peak $n_{\rm eff}$ around $10^3\,{\rm cm^{-3}}$. This enhancement is a 30x over the corresponding raw $n_H$ distribution. Coagulation activity declines steadily after cosmic noon and ceases entirely by $z\approx0.4$ as the dense ISM phase becomes increasingly sparse with declining star formation, consistent with
the high effective densities ($\gtrsim100\,n_H\,\rm{cm^{-3}}$,
\citealt{zhukovska08, asano13}) required to trigger grain--grain
collisions. The grain population built by this process was therefore established in earlier times and subsequently modified only by the destruction and growth channels that remain active at later times, rather than by ongoing coagulation itself.

Naturally the average grain radius increases ($+\sim 20\%$), but another less expected effect is that the D/G ratio drops considerably (-$\sim 36\%$). The larger dust grains lead to faster conversion of dust to gas (lowering the ratio) because they erode mass faster than smaller grains ($m \alpha a^2$). The result is that overall dust destruction slows, but mass loss speeds up.

If one reduces coagulation (by increasing parameter DustCoagulationCalibration (see Table \ref{tab:new_gadget4_params}) then this will reduce the average grain size, overall dust mass, and the D/G and D/Z within the ISM and $R_{200}$.

\textbf{Shattering:} S9 sees a drop in dust mass as particles are destroyed
once a shattering event drives their radius below the 1\,nm minimum.
Shattering requires the local turbulent velocity, derived from the sound speed, to exceed a composition-dependent threshold rising linearly from
$1$\,km/s for silicate to $2$\,km/s for carbon \citep{jones94}, and is
further restricted to diffuse gas, as coagulation occurs instead in dense gas. Once both gates are cleared, whether a given grain shatters on a given timestep is governed by a stochastic Poisson check utilizing the HK11 timescale. Full grain destructions are correspondingly rare, totaling $38{,}035$ events across the entire run. 

This destruction is overwhelmingly silicate ($38{,}018$ events) rather than
carbon ($17$ events), summing to $4.0\times10^8\,M_\odot$ of cumulative dust mass returned to the gas phase over the full simulation history. This dust mass is comparable to the entire z=0 dust mass, showing that most of this mass has been recycled through subsequent grain growth and creation processes. With the population of SNII-silicate ($10$\,nm) and AGB-carbon ($100$\,nm) grains, the silicate-heavy shattering destruction makes sense  as many fewer shattering events are needed to completely destroy the smaller grain. In addition, carbon grains higher shattering threshold favors their preferential survival even more strongly. This selection effect removes the smallest grains from the population and returns their mass to the gas phase, raising both the mass-weighted mean grain radius (+10\%) and the carbon fraction (+20\%).

\textbf{Radiation pressure:} S10 has relatively minor impacts on dust properties at this resolution. The instantaneous radiation-pressure acceleration is actually largest for the smallest grains (it scales as $Q_{\rm pr}(a)/a$, and $Q_{\rm pr}$ saturates well before $a$ does), but small grains are also the most tightly drag-coupled to the gas
(stopping time $t_{\rm stop}\propto a$), so this kick is damped almost immediately. In the terminal-velocity-limited regime relevant for sustained transport, the $a$-dependence of acceleration and of $t_{\rm stop}$ nearly cancel, leaving the terminal drift velocity set primarily by $Q_{\rm pr}(a)$ alone -- which still rises monotonically with grain size. Larger grains, particularly near our ceiling of
200\,nm, are therefore the most effectively transported and are deposited systematically into the CGM, which might affect observers of dust extinction curves or dust emission. 

\begin{figure}
    \includegraphics[width=0.47\textwidth]{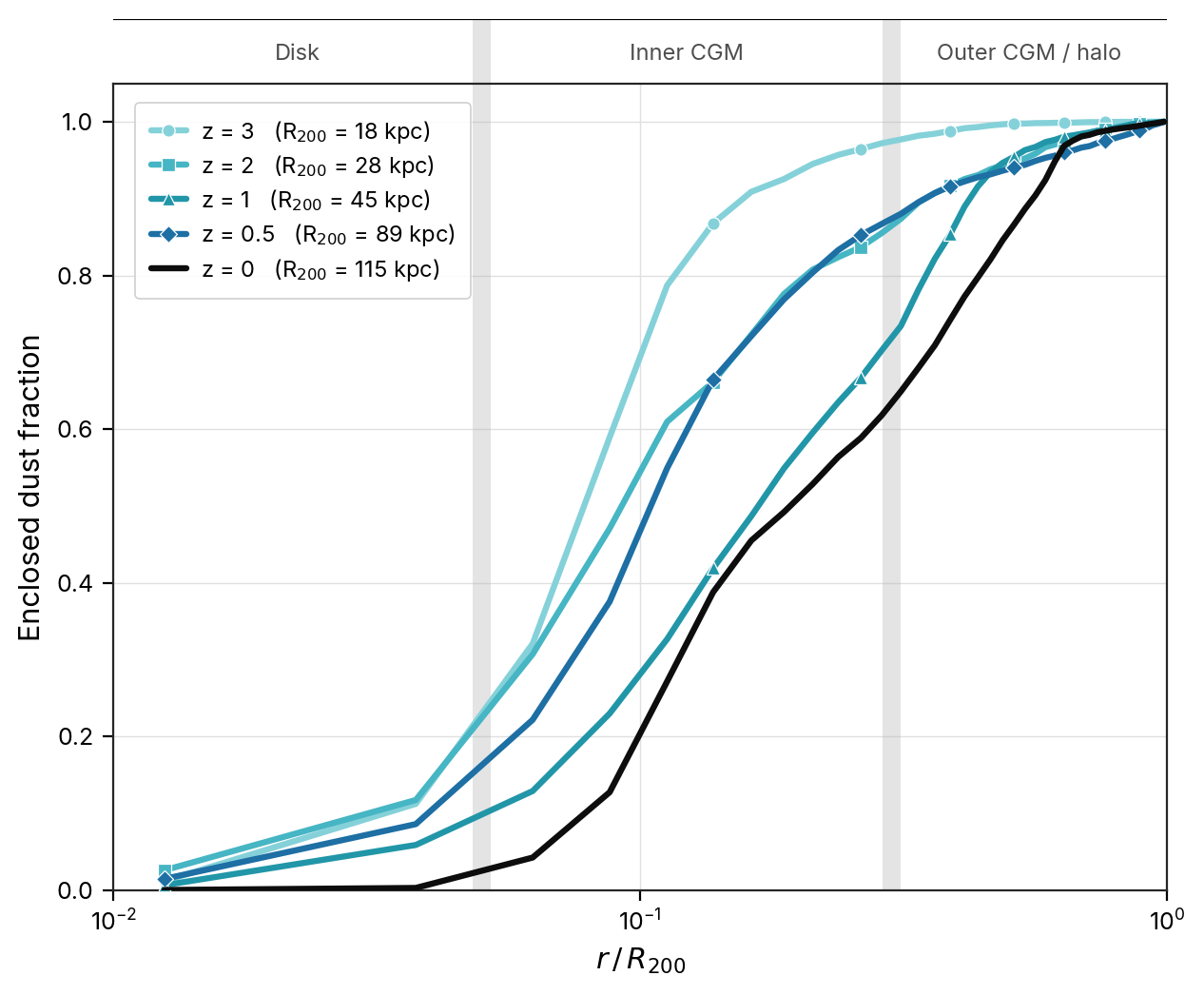}
    \caption{ 
    The cumulative radial distribution of dust for Halo 569 out to $R_{200}$ at five different redshifts in the CosmicGrain S10 $1024^3$ simulation. Curves show the enclosed fraction of the total dust mass as a function of normalized galactocentric radius (r/R200). Dust becomes progressively more extended with cosmic time as a growing fraction leave the stellar disk and reach the halo.
    }
    \label{fig:radial_dust_distribution}
\end{figure}

Figure~\ref{fig:radial_dust_distribution} confirms this transport mechanism directly: the cumulative enclosed dust fraction, shown at $z=4,2,1,0.5$, and $0$, becomes steadily more extended toward lower redshift, tracing the gradual relocation of dust from the disk into the CGM. The fraction of dust enclosed within the disk ($r/R_{200}\lesssim0.03$) drops from $z=1$ to $z=0$, consistent with sustained outward transport over several Gyr rather than a single rapid event. The progression is not perfectly monotonic between $z=2$ and $z=1$, which may reflect the documented major merger in Halo 569's assembly history.

Whether radiation pressure's impact strengthens at higher resolution is unclear. Resolving denser ISM structure could shorten grain-gas coupling timescales and suppress the kinetic kick, but also simultaneously place dust closer to individual star particles and increase the stellar flux. Distinguishing these potentially competing effects would require a higher resolution study which we defer to future work.

\begin{figure*}
    \includegraphics[width=0.99\textwidth]{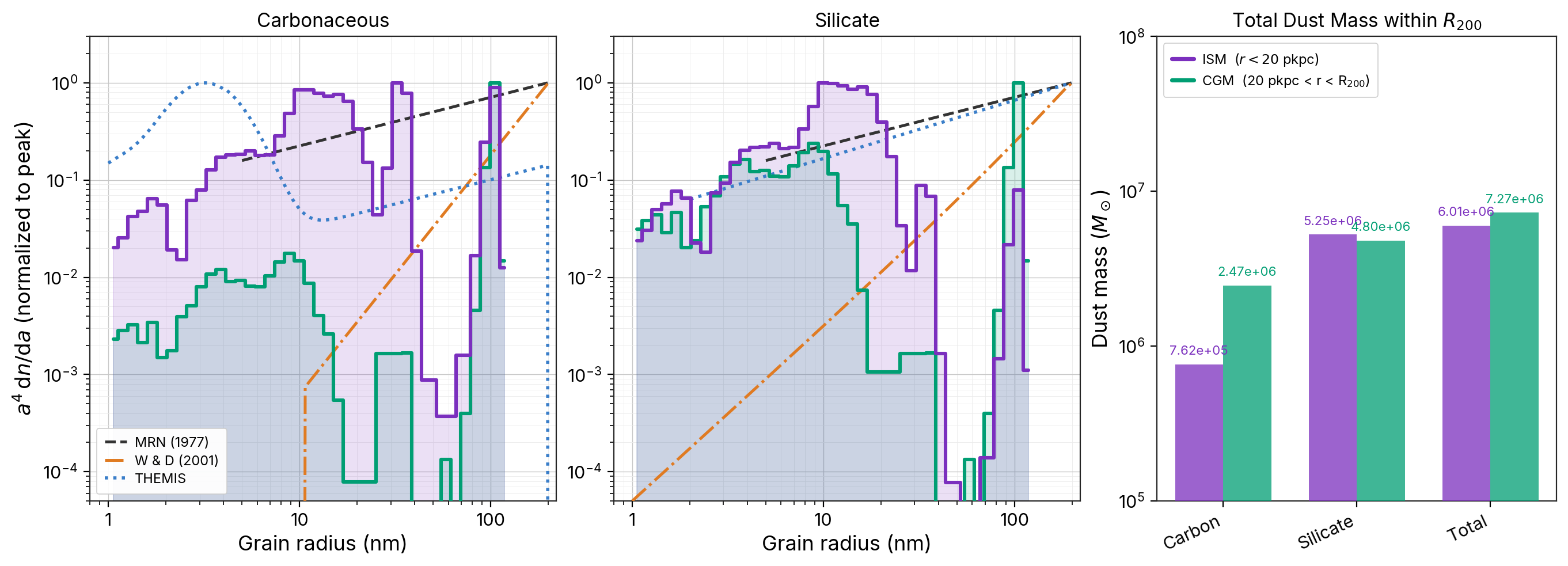}
    \caption{
Grain size distributions of the CosmicGrain S10 ($1024^3$) simulation within Halo 569 ($R_{200}$ = 113 pkpc at z=0) in the ISM ($r<20$ pkpc; purple) and CGM ($20~{\rm pkpc}<r<R_{200}$; green), compared to the classic power law (\citealt{mathis77}, MRN), the Milky Way dust model of \citet{weingartner01}, and the THEMIS dust model \citep{jones13,jones17}. Dust particles are separated into carbonaceous (left) and silicate (middle) populations according to their carbon fraction. The plotted distributions show ($a^4 dn/da$), a standard representation that emphasizes the distribution of dust mass across logarithmic size intervals. All grain size distributions are independently normalized to their peak values to facilitate comparison of their shapes. The right panel shows the corresponding integrated dust masses within the ISM and CGM, separated into carbonaceous, silicate, and total dust components.
    }
    \label{fig:grain_size_distribution}
\end{figure*}

\subsection{Dust Properties}

\begin{figure}
\includegraphics[width=0.47\textwidth]{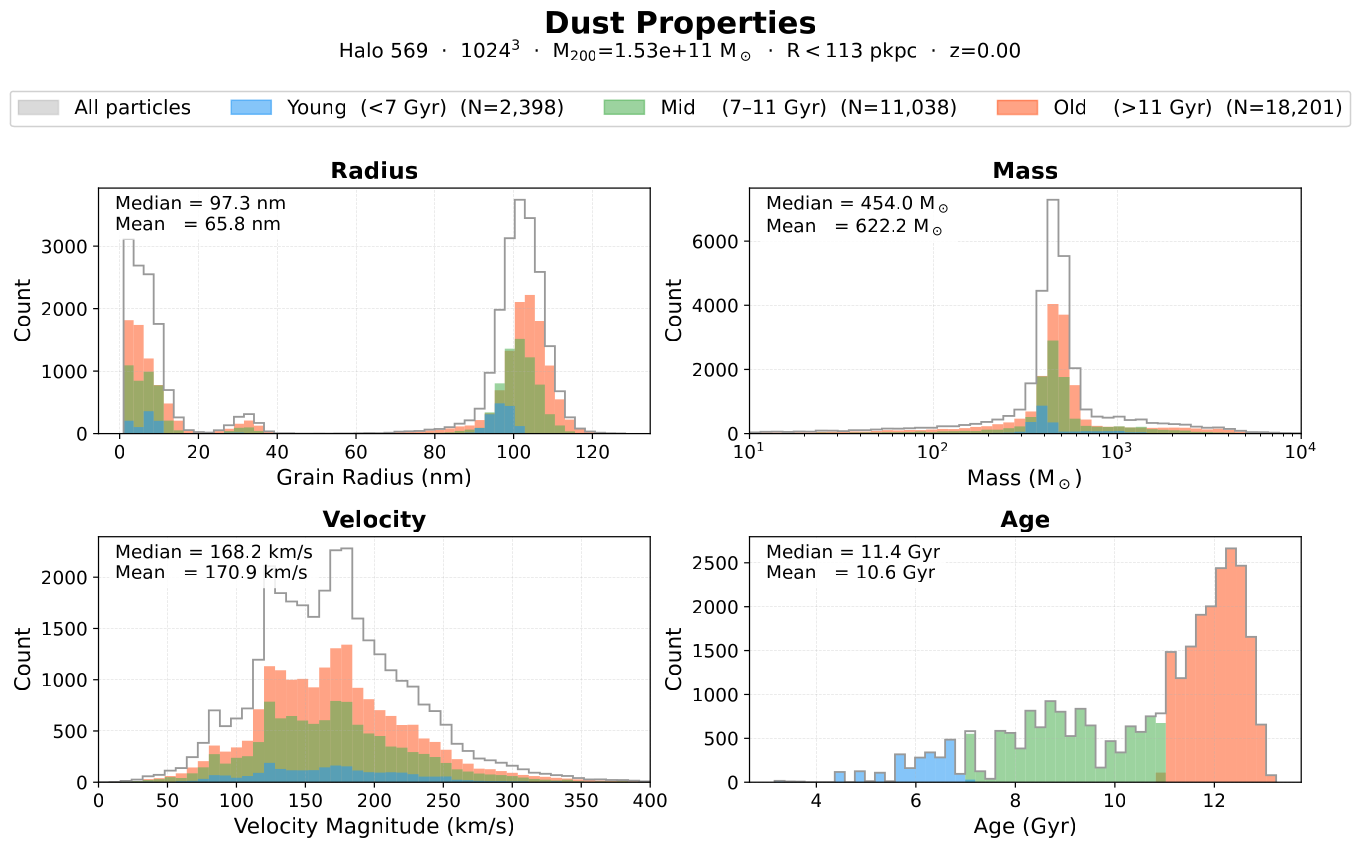}
    \caption{ 
    Histograms of dust properties (grain radius, mass, velocity, and age) color-coded by the dust particle's age (<7 Gyr is blue, 7-11 Gyr is green, >11 Gyr is salmon) within $R_{200}$ of Halo 569 at z=0 for the S10 $1024^3$ run. Grey outline represents the sum of all dust particles.
    }
    \label{fig:dust_properties}
\end{figure}

Figure~\ref{fig:dust_properties} shows age-coded histograms of key dust properties at $z=0$ within $R_{200}$ of Halo~569. The grain radius distribution is bimodal but dominated by number by the large-grain population. A minority small-grain population peaks near $\sim5$--$10$\,nm, corresponding to SNII-generated dust, while the dominant peak sits near $\sim100$\,nm (median $97$\,nm), reflecting AGB-generated grains which are subsequently grown by grain growth and coagulation. 

The mass distribution peaks near the median value of
$\sim450\,M_\odot$, reflecting the AGB creation mass
($\sim434\,M_\odot$ at $1024^3$), with a tail extending to
$\sim10^4\,M_\odot$ from grain growth and coagulation, reaching up to $\sim10\times$ the initial mass for the most enlarged surviving grains by $z=0$. 

The velocity distribution peaks near $\sim175$\,km/s with a tail racing out to $\sim350$\,km/s. This reflects the combination of SN-feedback kick velocities at grain birth and subsequent aerodynamic drag. Grains born in or near the dense ISM are decelerated efficiently, while grains that reach the more diffuse CGM retain larger velocities owing to longer stopping times there (Section~\ref{sec:drag}). The high-velocity tail traces grains launched onto more ballistic trajectories before substantial drag deceleration can occur.

The age diagnostic of the velocities reveals that grains of all ages -- young (<7 Gyr), intermediate (7--11 Gyr), and old (>11 Gyr) -- span essentially the same velocity range with no systematic offset toward lower velocities for the oldest population. If drag were continuously decelerating grains
over their lifetimes, old grains would be measurably slower than young ones by $z=0$. The absence of this trend indicates that dust velocities equilibrate to their local gas environment on timescales much shorter than a Gyr -- consistent with the ISM stopping times of $\sim$Myr implied by our drag implementation -- so that current environment, not birth time, governs the velocity a grain carries today.

The dust age distribution rises steeply toward $\sim11$--$13$\,Gyr (median $11.4$\,Gyr), reflecting the long-lived AGB population that dominates the $z=0$ grain population by being larger at birth despite carbonaceous grains being more sputter resistant (Section~\ref{sec:sputtering}). The near absence of young ($<7$\,Gyr) grains indicates that recent star formation contributes little to the surviving dust population, consistent with the declining star formation rate after cosmic noon and with newly formed grains being efficiently destroyed by thermal sputtering before reaching a size large enough to resist further erosion.

Figure~\ref{fig:grain_size_distribution} presents
number-weighted grain size distributions for ISM ($r<20$\,pkpc) and CGM ($20\,\mathrm{pkpc}<r<R_{200}$) dust at $z=0$, split by composition where all curves are independently normalized to their peak values so that slopes and shapes can be compared directly. The simulation produces distinct ISM and CGM grain size distributions, with the ISM showing a relative enhancement of large grains. Despite occupying a much larger volume, the CGM contains a total dust mass comparable to that of the ISM within $R_{200}$, indicating that a substantial fraction of the dust produced by stars survives relocation into the halo and lingers in the circumgalactic environment at $z=0$. While both regions end up with similar amounts of silicate dust, the CGM retains nearly $\sim3\times$ more carbon dust. This carbon
preference traces directly to the combination of coagulation and
gas-dust coupling as larger AGB-sourced carbon grains have longer
stopping times in the dense ISM, allowing them to travel further before re-coupling, which preferentially relocates carbon into the CGM relative to the smaller more tightly-coupled silicate population. This may have some bearing on the ongoing difficulty cosmological simulations have reproducing sightline observations of carbon abundances \citep{rahmati16, finlator20}.

Interestingly, FIRE simulations that begin with continuous grain size distributions also develop distinct small and large-grain populations through the combined effects of shattering, coagulation, and grain growth \citep[][their Figure 6]{choban26}. \CG exhibits a qualitatively similar two-component grain population, although in our model the characteristic grain sizes are initially seeded at specific radii by SNII ($10$\,nm) and AGB ($100$\,nm) dust sources and subsequently modified by grain processing. Starting from this initial bimodality, the broader distributions seen in the full-physics S10 model show that subsequent grain processing substantially reshapes
the dust population though does not erase the bimodal distribution entirely.

\begin{figure}
    \includegraphics[width=0.49\textwidth]{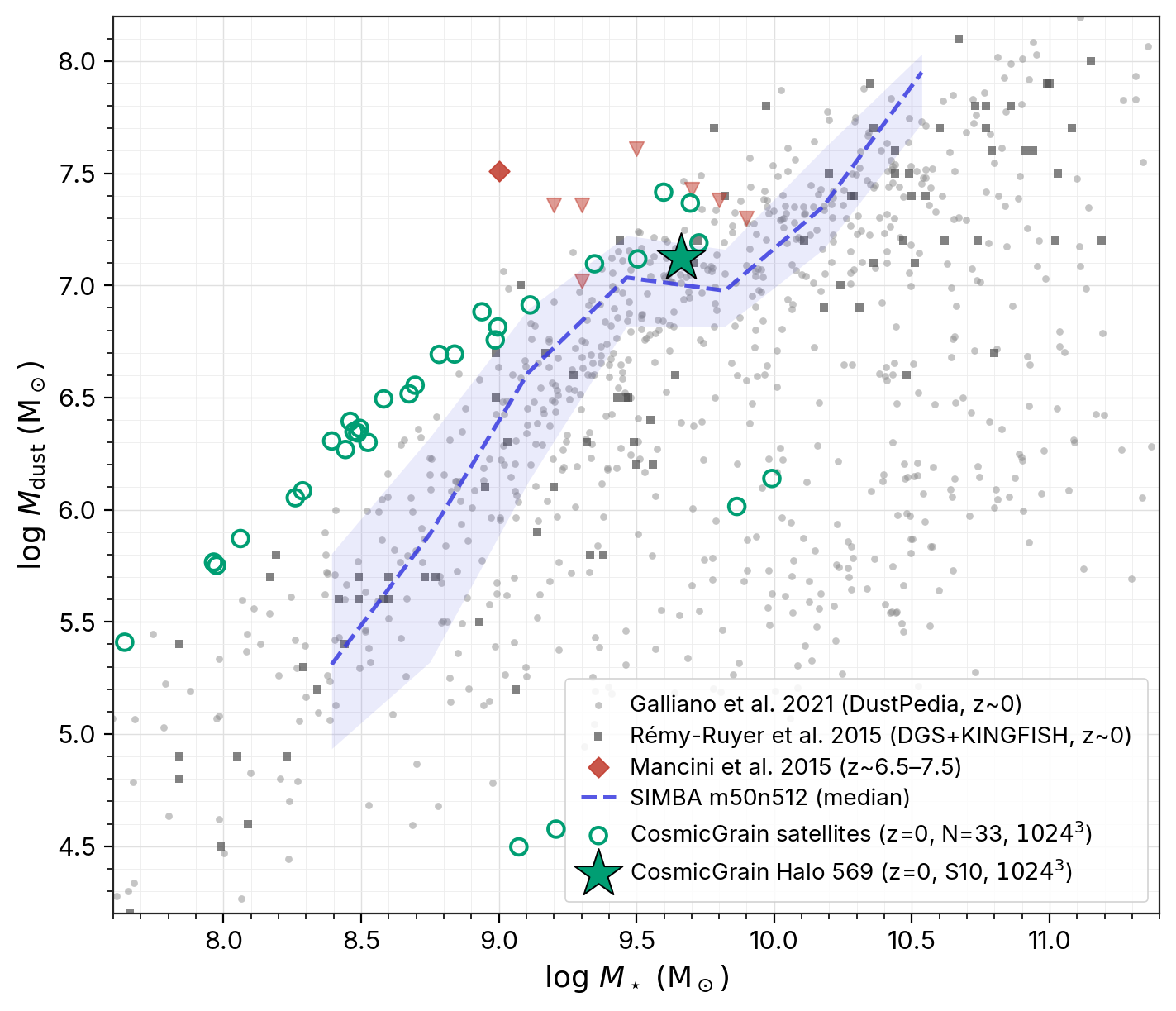}
    \caption{ 
    A comparison of the stellar mass to dust mass between our primary zoom halo (569), 34 other smaller halos within our zoom region, $\sim 800$ galaxies from the DustPedia survey \citep{galliano21}, 109 galaxies from DGS and KINGFISH \citep{rr15}, and the single detection and additional upper limits of dusty high -z galaxies from \citealt{mancini15}, as well as median binned values from a 50 Mpc SIMBA simulation \citep{dave19}. The SIMBA medians retain only galaxies with ongoing star formation ($sSFR > 1/t_H(z)$).
    }
    \label{fig:stellar_vs_dust_mass}
\end{figure}

\subsection{Comparison to Observations}

The growth of Halo 569's total dust mass with stellar mass follows the rising slope of THESAN \citep[][their Fig.\ 15a]{k22} and falls neatly within ALMA observations at $z\sim6$ \citep{mancini15}.  Figure~\ref{fig:stellar_vs_dust_mass} plots this evolution for our
primary zoom halo and 34 smaller halos in the high-resolution region, alongside the DustPedia survey ($\sim800$ galaxies) and the combined DGS and KINGFISH surveys (109 galaxies). Halo 569 (green star) sits about 0.5\,dex above the SIMBA median and the bulk of the observed sample at its stellar mass ($\log M_\star = 9.8$). While \CG predicts more dust than is typical in SIMBA or the large surveys, the \citet{mancini15} detection (red diamond) and upper limits (triangles) show where dusty high-redshift galaxies fall and these match our results well at $\log M_\star = 9.5$--$10$.

The final row of Table~\ref{tab:dust_sim_ladder_1024} lists Milky Way values from the literature for comparison, including $M_\star/M_{\rm halo}$ computed from $M_\star=5\times10^{10}\,M_\odot$ \citep{blandhawthorn16} and
$M_{\rm halo}=1.3\times10^{12}\,M_\odot$ \citep{watkins19}; $D/G$ and
$M_{\rm dust}/M_\star$ from \citet{draine07}; and $D/Z$ as the median
of 37 Galactic ISM sight lines from Table~B.2 of \citet{konstantopoulou24}. $D/G|_{\rm ISM}$ and $D/Z|_{\rm ISM}$ equal the corresponding global values because these measurements sample individual ISM sight lines exclusively, with no CGM contribution. Halo 569 has an order of magnitude more dust mass than is measured in the Milky Way \citep{draine07} which we attribute to the ineffectiveness of the current supernova shock destruction treatment (see Section \ref{sec:dust_physics}) and of our model's lack of AGN feedback which may lead under-predicting dust expulsion.

\begin{figure}
    \includegraphics[width=0.48\textwidth]{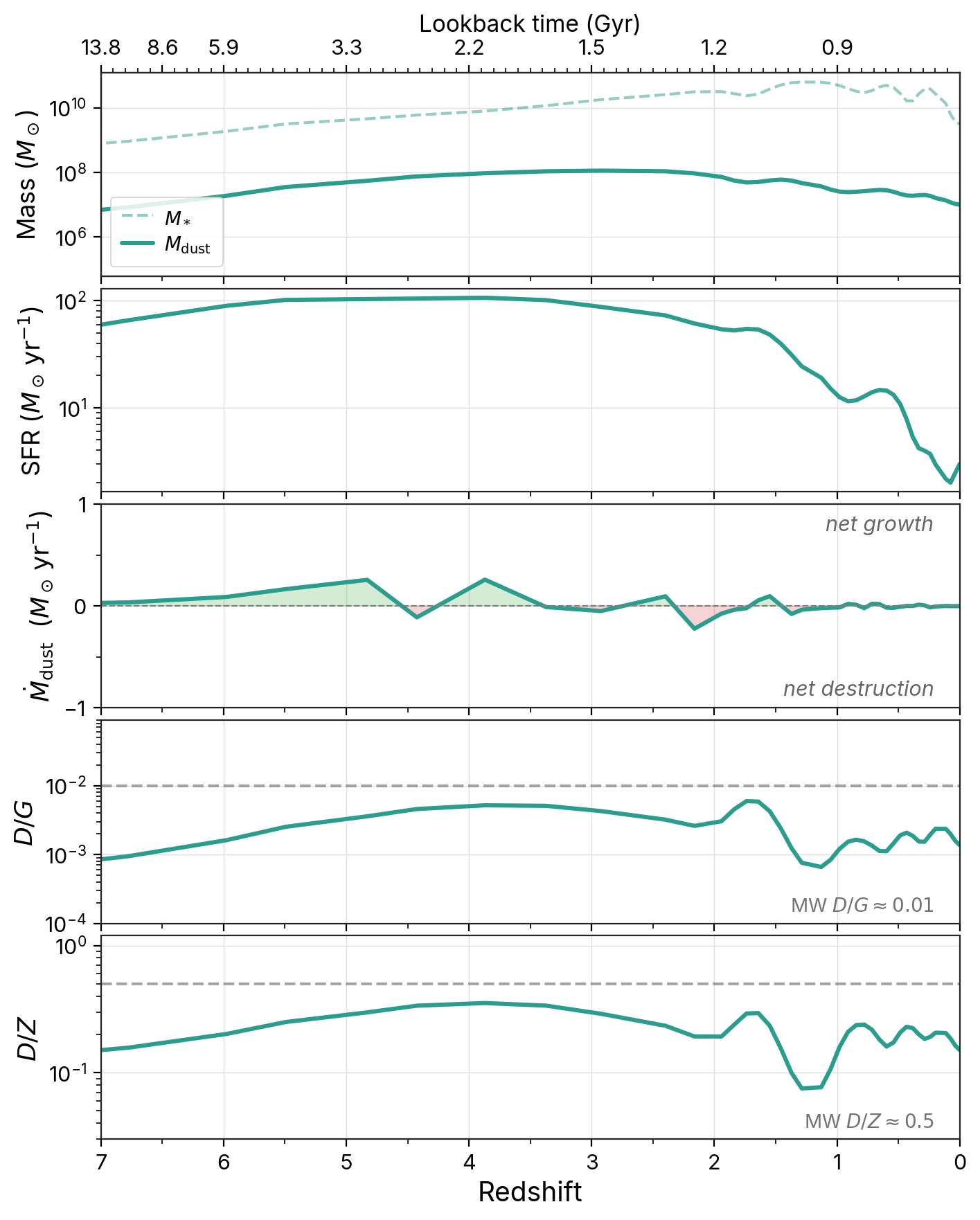}
\caption{Cosmic evolution of Halo 569 (S10, $1024^3$) within $R_{200}$. \textit{Top:} stellar mass (dashed) and dust mass (solid) vs.\ redshift. \textit{Second:} star formation rate. \textit{Third:} net rate of change of dust mass, $\dot{M}_\mathrm{dust}$, shaded green where dust mass is growing and red where it is being destroyed. \textit{Fourth and fifth:} dust-to-gas ($D/G$) and dust-to-metals ($D/Z$) ratios, with Milky Way values \citep{draine07,konstantopoulou24} shown as dashed grey lines for reference. The corresponding lookback time is given along the top axis. Short-timescale fluctuations at $z\lesssim2$, most apparent in $\dot{M}_\mathrm{dust}$, $D/G$, and
$D/Z$, coincide with a period of substructure in the halo's assembly history and should not be over-interpreted as intrinsic physical evolution.
    }
    \label{fig:evolution_comparison}
\end{figure}

The cosmic evolution in Halo 569 of stellar and dust mass, star formation rate, the rate of change of dust mass, and the D/G and D/Z ratios are shown in Figure \ref{fig:evolution_comparison}. Observations in sub-millimeter emission have shown that there is a delay of $\sim 1$ Gyr between the peak of star formation and the peak of dust abundance \citep{parente25}. Our S10 model finds that peak dust occurs at $z\sim2.4$, and the peak of star formation at $z\sim3.4$, suggesting a qualitatively similar but somewhat smaller delay of $\sim0.85$ Gyr.

\begin{figure}
    \includegraphics[width=0.47\textwidth]{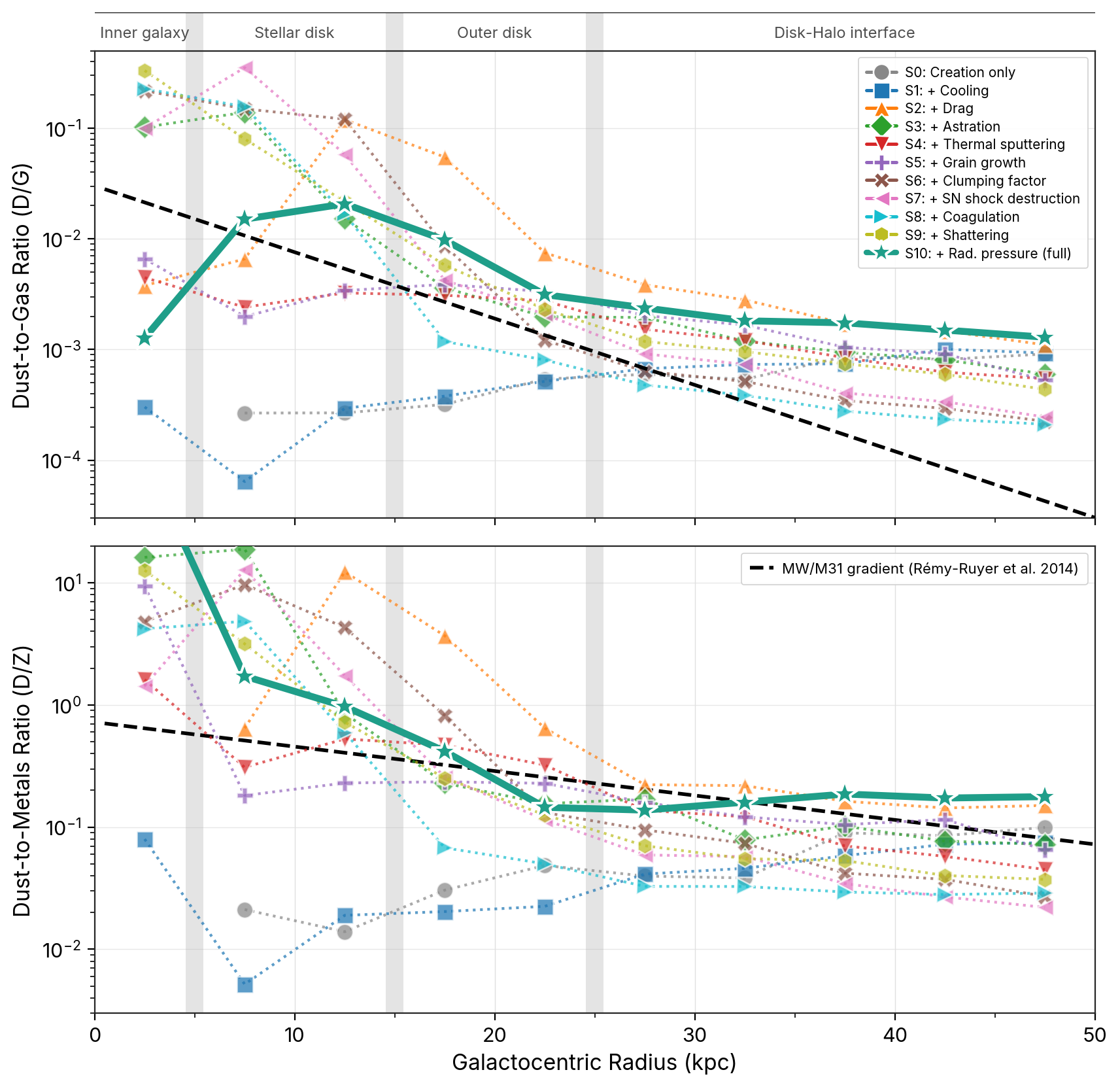}
    \caption{ 
Dust-to-Gas and Dust-to-Metal ratios as a function of galactocentric radius at redshift z=0 for the $1024^3$ CosmicGrain physics ladder. Profiles are measured in 5 pkpc spherical shells centered on Halo 569 and extend to 40 pkpc. Colored curves show the cumulative addition of dust physics processes from the creation-only model (S0) to the full-physics model (S10). Dashed reference curves are constructed from the observed D/G–metallicity relation of \citet{remyruyer14}, assuming a typical spiral-galaxy metallicity gradient of -0.04 dex kpc$^{-1}$ \citep{lemasle18}. S10 produces radial gradients broadly consistent with local spiral galaxy observations.
    }
    \label{fig:dust_ratios}
\end{figure}

The radial $D/G$ and $D/Z$ profiles of Halo 569 are shown in
Figure~\ref{fig:dust_ratios}, highlighting the impact of individual dust processes on the shape of each profile. Without drag included, S0 and S1 have almost no dust remaining in the ISM at z=0, though they converge with the other runs by $\sim40$\,kpc. Beyond this radius, individual dust physics have
minimal effect, with $D/G$ converging on $10^{-3}$ and $D/Z$ on $10^{-1}$ across all runs. The full-physics S10 model agrees reasonably
well with observational expectations out to the disk-halo interface
($\lesssim40$\,kpc) \citep{remyruyer14}, but predicts a much flatter
$D/G$ and $D/Z$ gradient at larger radii, reflecting dust transport
beyond the outer disk and and into the CGM.

\begin{figure}
    \includegraphics[width=0.47\textwidth]{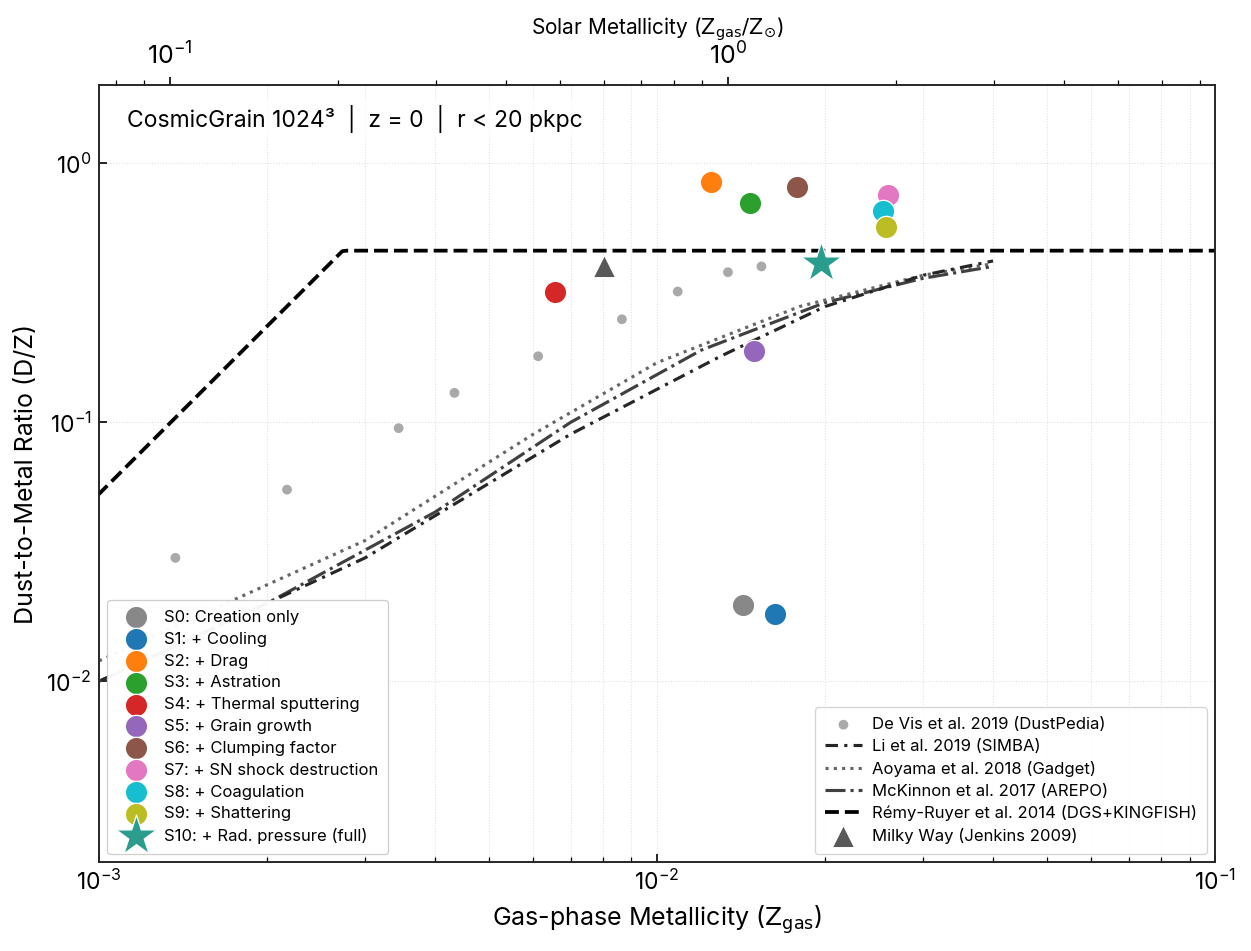}
    \caption{ 
Dust-to-metal ratio as a function of gas-phase metallicity for the CosmicGrain physics ladder (S0--S10) at z=0. Quantities are measured within a fixed 20 pkpc radius centered on Halo 569. The gas-phase metallicity on the x-axis is computed as a mass-weighted metallicity of the enclosed gas, while the dust-to-metal ratio on the y-axis is defined as the total dust mass divided by the total metal mass within the same region. Colored symbols show individual CosmicGrain runs, progressing from dust creation only (S0) to the full-physics model (S10). Observational constraints from \citet{remyruyer14}, \citet{devis19}, and Milky Way measurements are shown for comparison, along with results from the dust evolution models of \citet{mckinnon17}, \citet{aoyama18}, and \citet{li19}.
    }
    \label{fig:dz_vs_zgas}
\end{figure}

Figure~\ref{fig:dz_vs_zgas} shows $D/Z$ as a function of gas-phase
metallicity within the inner 20\,pkpc, compared against three
observational and three modeled studies. The complete model (S10, teal
star) falls squarely within the observational surveys of DGS and
KINGFISH \citep{remyruyer14} and the trend of DustPedia's Model~VI
\citep[][their Figure 10]{devis19}, as well as the SIMBA \citep{li19} and AREPO \citep{mckinnon17} models, giving us confidence that S10 produces a realistic $D/Z$ at high metallicity. The ladder's evolution along this relation illustrates the importance of balancing all dust physics together: models including only a subset of the processes generally produce dust-to-metal ratios well above the observed relation, while the full-physics model converges toward observational expectations. No single process dominates dust evolution; rather, the observed $D/Z$ relation emerges from the interplay between dust production, grain growth, and destruction operating together within the ISM.

\subsection{Resolution Convergence}

In ideal circumstances, increasing the resolution of a simulation would result in unique star formation histories and particle structure, but still yield identical results, such as in total dust mass or feedback efficiency. Grain growth, however, is an example of a density-dependent process that will be accelerated by a greater simulation resolution which resolves higher density and colder gas clouds resulting in more dust mass overall.

We minimize resolution-dependent parameter changes where possible (Table~\ref{tab:new_gadget4_params} lists the few that are adjusted, including \texttt{DustParticlesPerSNII/AGB}, tuned to hold $M_{\rm dust}/M_{\rm gas}$ fixed as per-particle mass decreases with resolution; see Methods). We adopt $1024^3$ as the primary resolution for the physics ladder (Section~\ref{tab:dust_sim_ladder_1024}), with
$2048^3$ serving as a convergence check and $512^3$ included for completeness only -- at this resolution, sparse ISM dust produces an anomalously large $\langle a\rangle_{\rm ISM}$ that reflects sampling, not real grain growth.

With the Jeans pressure floor applied (Section~\ref{sec:simulation}),
the $2048^3$ S10 run recovers a stellar mass fraction, ISM grain size, and dust-to-gas ratio broadly comparable to $1024^3$ (Table~\ref{tab:dust_sim_ladder_1024}). $M_\star/M_{\rm halo}$ and the halo-wide mean grain size $\langle a\rangle_{200}$ are well converged between the two resolutions, agreeing to within $5\%$, indicating the global grain size distribution is robustly captured already at $1024^3$. The ISM-aperture grain size $\langle a\rangle_{\rm ISM}$ decreases modestly with resolution, consistent with thermal sputtering being better resolved at higher particle counts. The global dust budget is more resolution-sensitive: both $D/G$ and $M_{\rm dust}/M_\star$ decline from $1024^3$ to $2048^3$, consistent with destruction processes -- thermal sputtering and SN shock destruction in particular -- becoming more effective once feedback energy is deposited into lower-mass particles.

\subsection{Future Improvements}

In nature, drag forces between gas and dust is a two-way coupling. Momentum conservation dictates that gas affects the movement of dust particles, and dust particles should conversely push gas particles around. In our initial codebase here, we implement the former half only as we expect this back-reaction to be small outside the realm of our focus on dust processes.

The treatment for dust destruction from shocks assumes spherical symmetry, uniform ambient density, no energy losses from radiation, and no magnetic fields or cosmic rays. All of this additional physics would improve the accuracy of dust on an evolving halo. In addition, the shock radius can become too large at late times and over-dilute shocked dust particles making the supernovae destruction inefficient. A revised interaction with the spatial hashes will be implemented in the next code update.

Supernovae shocks and subsequent dust destruction are affected by dense cold clumps of gas because of changes in the propagation speed and possibly due to shielding \citep{asano13, mckinnon17}. The subgrid management process of clumping (see \ref{sec:clumping}) which benefits grain growth and coagulation would also improve a shock destruction treatment though we do not include this here.

We have introduced a constant density factor representing the transition from a regime that coagulates dust grains to one that shatters them. This is overly simplified, as it depends on grain size, velocity dispersion, and dust-to-gas ratio simultaneously. Density is used a proxy for velocity, but an explicit velocity-based would improve the model. The literature suggests that densities $\geq 100 n_H cm^-3$ are required to begin coagulation \citep{zhukovska08, asano13}. Even with the clumping technique bridging the gap in resolved densities, the zoom resolutions $\geq 1024^3$ do not reach the densities needed for dust particles to coagulate.  

The transition between coagulation and shattering regimes is governed
by a fixed density threshold (\texttt{DustCollisionDensityThresh}),
augmented by a subgrid clumping factor to account for unresolved dense
structure. In our simplified implementation, gas density serves as a proxy for the turbulent velocity dispersion driving grain--grain collisions, since denser environments see higher velocity dispersions, rather than tracking collision
velocities explicitly. A better treatment would replace this
proxy with a subgrid turbulent velocity dispersion estimated from the
local gas velocity field, better capturing coagulation and shattering
in environments where density and turbulent energy diverge, such as
post-shock gas or the outer CGM.


The molecular fractions which inform grain growth efficiency we use currently ($f_{mol}$; see Section \ref{sec:growth}) are generalized density-dependent values. More sophisticated molecular fraction models like those of \citealt{krumholz11} would upgrade this approach. Additionally, we use a generic metal reservoir to gain and remove metallicity to and from gas and dust. Element-specific pathways as shown by \citealt{zhukovska08} are left for future work.

Radiation pressure currently relies on a simplified age-based luminosity relation restricted to the youngest (< 20 Myr) stellar populations. Future work will replace this with UV luminosities from a population synthesis model such as BPASS \citep{eldridge2017}, interpolated by age and metallicity, extending the treatment to a wider range of stellar ages while remaining consistent with the UV-calibrated grain absorption efficiency already assumed in the model.

\section{Summary}
\label{sec:conclusions}

We have developed a new dust model in the smoothed particle hydrodynamics code \gadget, accounting for local stellar production of dust from SNe and AGB stars, growth in the ISM due to accretion of gas-phase metals, destruction via SN shocks, thermal sputtering, and astration. Dust particles are active on-the-fly particles which individually  experience drag and radiation pressure. We run a zoom simulation focused on one Milky Way mass halo (Halo 569), and isolate each individual dust process in sequence labeled S0 to S10.

We summarize several takeaways:

\begin{enumerate}
\item Modeling dust as active, on-the-fly superparticles allows individual grains to retain distinct formation histories, enabling results inaccessible to passive dust models.

\item The silicate-to-carbon shattering destruction ratio is strongly asymmetric by roughly three orders of magnitude, arising because silicate grains are born smaller and reach the destruction size threshold after fewer collisions
than their carbon counterparts, a result that requires individual grain history to deduce.

\item The $z=0$ grain velocity distribution is broad and
age-independent. Grains formed across more than 10 Gyr of cosmic history share essentially the same kinematic spread, indicating that current local environment influences dust's present-day velocity more than its birth conditions.

\item A substantial fraction of carbon dust mass resides
in the CGM rather than the ISM at $z=0$. Silicate dust remains comparably split between the two environments.

\item Gas-dust drag is the dominant driver of ISM dust enrichment, increasing dust levels by $\sim 100\times$. All other physics channels affect the R200 dust budget by only $10$--$30\%$, underscoring drag as the primary regulator of where dust resides rather than how much exists.

\item Shock destruction proved the most challenging dust physics to implement as its relevant length scale ($\sim 50$pc) lies far below what our spatial-hash neighbor-finding can resolve at the zoom-region scale. Independent of this, grain growth (amplified by subgrid density clumping) is the important mass-building channel at $z\gtrsim2$, while thermal sputtering emerges as the dominant destruction mechanism at $z\lesssim2$.

\item Overall dust mass follows the star formation history with a several hundred Myr delay between the peak of star formation and the peak of dust mass, though the precise value is sensitive to short-timescale fluctuations in this halo's assembly history.

\item Global quantities of $M_\star/M_{\rm halo}$ and halo-wide mean grain size are well converged between $1024^3$ and $2048^3$ resolutions, while the total dust budget is more resolution-sensitive.
\end{enumerate}

\section{Data Availability}
\label{sec:data}
The complete \CG code is freely available at \url{https://cosmicgrain.space} under
the GNU General Public License v3.0 (GPLv3), and is archived on Zenodo at \url{https://doi.org/10.5281/zenodo.21302432} corresponding to the version used to produce the results in this paper. Example runs, analysis scripts, and initial conditions files are also included. The full simulation snapshots produced for this work are available upon request.

\software{Gadget-4 \citep{springel21},
          CAMB \citep{lewischallinor11},
          MUSIC2 \citep{hahn11},
          NumPy \citep{harris20},
          SciPy \citep{virtanen20},
          Matplotlib \citep{hunter07},
          h5py \citep{collette14}
          }

\begin{acknowledgments}
EH would like to thank Harrison Cook and Andrew Saydjari for insightful discussions and technical support that greatly benefited this work. EH is supported by the NASA FINESST program, award \#80NSSC24K1483. This work utilized computational resources from the New Mexico State University High Performance Computing Group, which is directly supported by the National Science Foundation (OAC-2019000), and from the Stampede3 system at the Texas Advanced Computing Center (TACC) at The University of Texas at Austin through allocation PHY250353 from the Advanced Cyberinfrastructure Coordination Ecosystem: Services and Support (ACCESS) program, which is supported by National Science Foundation grants 2138259, 2138286, 2138307, 2137603, and 2138296.
Stampede3 is funded by the National Science Foundation (OAC-2320757). KF and EH gratefully acknowledge support from STScI Program \#HST-AR-16125.001-A. The Cosmic Dawn Center is funded by the Danish National Research Foundation. Portions of this project were improved with the AI assistance of Claude (Anthropic), ChatGPT (OpenAI), and MS Visual Studio (Microsoft). All scientific content, numerical results, and conclusions were verified by the authors.
\end{acknowledgments}

\begin{contribution}
\textbf{EH}: Conceptualization, Investigation, Software, Visualization, Writing - Original draft; \textbf{KF}: Conceptualization, Supervision, Writing - Review \& Editing.
\end{contribution}

\clearpage
\appendix

\renewcommand{\thetable}{A\arabic{table}}
\setcounter{table}{0}

\begin{table}[t]
\centering
\caption{CosmicGrain Dust Model Parameters. Parameters with multiple values are resolution-dependent, shown as $512^3$ / $1024^3$ / $2048^3$. All other parameters are fixed across resolutions.}
\label{tab:new_gadget4_params}
\setlength{\tabcolsep}{5pt}
\renewcommand{\arraystretch}{1.15}
\small
\begin{tabular}{l l l}
\hline
\textbf{Parameter} & \textbf{Value} & \textbf{Description} \\
\hline
\multicolumn{3}{l}{\textit{General grain-size limits}} \\
\texttt{DustMinGrainSize} & $1.0\,\mathrm{nm}$ &
  Minimum grain radius \\
\texttt{DustMaxGrainSize} & $200\,\mathrm{nm}$ &
  Maximum grain radius \\
\hline
\multicolumn{3}{l}{\textit{Dust creation}} \\
\texttt{DustYieldSNII}$^{1}$ & 0.2 &
  Fraction of SNII metal mass condensing into dust \\
\texttt{DustYieldAGB}$^{2}$ & 0.4 &
  Fraction of AGB metal mass condensing into dust \\
\texttt{DustVelocitySNII} & $30\,\mathrm{km\,s^{-1}}$ &
  Radial kick velocity for newly spawned SNII dust \\
\texttt{DustVelocityAGB} & $10\,\mathrm{km\,s^{-1}}$ &
  Radial kick velocity for newly spawned AGB dust \\
\texttt{DustOffsetMinSNII} / \texttt{DustOffsetMaxSNII} & 0.02 / 0.10\,kpc &
  Minimum/maximum spawn offset from the parent star (SNII) \\
\texttt{DustOffsetMinAGB} / \texttt{DustOffsetMaxAGB} & 0.01 / 0.05\,kpc &
  Minimum/maximum spawn offset from the parent star (AGB) \\
\texttt{DustParticlesPerSNII} & 4 / 8 / 16 &
  Superparticles spawned per SNII event \\
\texttt{DustParticlesPerAGB} & 6 / 12 / 24 &
  Superparticles spawned per AGB event \\
\hline
\multicolumn{3}{l}{\textit{Grain destruction}} \\
\texttt{DustThermalSputteringTemp}$^{3}$ & $5\times10^5\,\mathrm{K}$ &
  Temperature threshold for thermal sputtering \\
\texttt{DustShockAmbientDensity}$^{5,6}$ & 0.1 / 0.5 / 2.0 &
  Ambient ISM density for Sedov-Taylor shock radius ($\mathrm{cm^{-3}}$) \\
\hline
\multicolumn{3}{l}{\textit{Grain growth and collisional processes}} \\
\texttt{DustCollisionDensityThresh}$^{4,8}$ & 15 / 100 / 150 &
  $n_\mathrm{eff}$ threshold separating coagulation and shattering
  regimes ($\mathrm{cm^{-3}}$) \\
\texttt{DustAccommodationCoeff}$^{9}$ & 0.3 &
  Gas-grain thermal coupling efficiency \\
\texttt{DustGrowthCalibration}$^{4}$ & 0.3 / 0.4 / 0.5 &
  Scaling factor for grain growth timescale \\
\texttt{DustShatteringCalibration}$^{8}$ & 1.0 / 1.2 / 0.7 &
  Scaling factor for shattering timescale \\
\texttt{DustCoagulationCalibration}$^{4}$ & 1.0 / 7.0 / 10.0 &
  Scaling factor for coagulation timescale \\
\hline
\multicolumn{3}{l}{\textit{Grain dynamics}} \\
\texttt{DustRadiationPressureEfficiency}$^{7}$ & 1.0 &
  Global scaling factor applied to the radiation pressure force \\
\hline
\multicolumn{3}{l}{\textit{Star formation (resolution-dependent)}} \\
\texttt{CritPhysDensity} & 0.1 / 0.7 / 1.0 &
  Star formation density threshold ($\mathrm{cm^{-3}}$) \\
\hline
\end{tabular}

\vspace{4pt}

\footnotesize
$^{1}$\citet{todini_ferrara01}; \quad
$^{2}$\citet{ferrarotti06}; \quad
$^{3}$\citet{mckinnon17}; \\
$^{4}$\citet{hk11}; \quad
$^{5}$\citet{sedov59}; \quad
$^{6}$\citet{bocchio14}; \\
$^{7}$\citet{draine84}; \quad
$^{8}$\citet{hirashita09}; \quad
$^{9}$\citet{hollenbach79}
\end{table}

\clearpage
\bibliography{citations}
\bibliographystyle{aasjournalv7}

\end{document}